\documentclass[review,3p,times]{elsarticle}

\usepackage[latin9]{inputenc}
\usepackage{bm}
\usepackage{amsmath}
\usepackage{amssymb}
\usepackage{graphicx}
\usepackage{natbib}
\biboptions{sort&compress}
\usepackage{tikz}
\usepackage{graphicx, psfrag}
\usetikzlibrary{calc,through}

\usepackage{epstopdf}

\usepackage{esint}
\usepackage{multirow}
\usepackage[subfigure]{graphfig}
\usepackage{rotating}
\usepackage{verbatim}
\makeatletter
\newcommand*{\rom}[1]{\expandafter\@slowromancap\romannumeral #1@}

\usepackage{tabularx}\usepackage{subfigure}\usepackage{subfigure}\usepackage{color}

\usepackage{bm}%% The

\journal{}

\makeatother

\color{black}

\graphicspath{}
\begin{document}
\title{A new paradigm of dissipation-controllable, multi-scale resolving schemes for compressible flows}

\author[ad1]{Xi Deng}

\author[ad2]{Zhen-hua Jiang \corref{cor}}

\author[ad1]{Peter Vincent}

\author[ad3]{Feng Xiao}

\author[ad2]{Chao Yan}

\address[ad1]{Department of Aeronautics,Imperial College London,SW7 2AZ, United Kingdom.}

\address[ad2]{College of Aeronautics Science and Engineering, Beijing University of Aeronautics and Astronautics, Beijing 100191, PR China.}

\address[ad3]{Department of Mechanical Engineering, Tokyo Institute of Technology, 2-12-1 Ookayama, Meguro-ku, Tokyo, Japan.}

\cortext[cor]{Corresponding author: Dr. Zhen-Hua Jiang (Email: jiangzhenhua@buaa.edu.cn)}

\begin{abstract}
The scale-resolving simulation of high speed compressible flow through direct numerical simulation (DNS) or large eddy simulation (LES) requires shock-capturing schemes to be more accurate for resolving broadband turbulence and robust for capturing strong shock waves. In this work, we develop a new paradigm of dissipation-controllable, shock capturing scheme to resolve multi-scale flow structures in high speed compressible flow. The new scheme employs a polynomial of n-degree and non-polynomial THINC (
Tangent of Hyperbola for INterface Capturing) functions of m-level steepness as reconstruction candidates. These reconstruction candidates are denoted as PnTm. From these candidates, the piecewise reconstruction function is selected through the boundary variation diminishing (BVD) algorithm. Unlike other shock-capturing techniques, the BVD algorithm effectively suppresses numerical oscillations without introducing excess numerical dissipation. Then, a controllable dissipation (CD) algorithm is designed for scale-resolving simulations. This novel paradigm of shock-capturing scheme is named as PnTm-BVD-CD. The proposed PnTm-BVD-CD scheme has following desirable properties. First, it can capture large-scale discontinuous structures such as strong shock waves without obvious non-physical oscillations while resolving sharp contact, material interface and shear layer. Secondly, the numerical dissipation property of PnTm-BVD-CD can be effectively controlled between n+1 order upwind-biased scheme and non-dissipative n+2 order central scheme through a simple tunable parameter $\lambda$. Thirdly, with $\lambda=0.5$ the scheme can recover to n+2 order non-dissipative central interpolation for smooth solution over all wavenumber, which is preferable for solving small-scale structures in DNS as well as resolvable-scale in explicit LES. Finally, the under-resolved small-scale can be solved with dissipation controllable algorithm through so-called implicit LES (ILES) approach. Through simulating benchmark tests involving multi-scale flow structures and comparing with other central-upwind schemes, the superiority of the proposed scheme is evident. Thus, this work provides an alternative scheme for solving multi-scale problems in high speed compressible flows.           

\end{abstract}

\maketitle

\section{Introduction}
Numerical simulation of compressible flow involving multi-scale flow features remains one of unsolved issues that are of great real life application importance. For example, in high speed compressible turbulent flow the interactions between shock waves and turbulence raise challenges to the design of numerical methods because the contradictory properties of numerical schemes are required in dealing with discontinuous large-scale and smooth small-scale features simultaneously. In nowadays community of computational fluid dynamics, the multi-scale flow features in compressible turbulence are resolved with so-called scale-resolving simulation through direct numerical simulation (DNS) or large eddy simulation (LES). The scale-resolving simulation requires high resolution numerical schemes which are low-dissipative to resolve small-scale structures in transitional and turbulent flows, and meanwhile are robust to stabilize solutions around large-scale features such as strong shocks, contact discontinuities and shear layers. A great deal of effort has been devoted to developing such numerical methods, and at present the shock-capturing schemes are the most popular methods among broad algorithms in the literature.

The shock-capturing methods can be classified as two types of quite different methods: the central schemes and the upwind schemes. The central schemes consider the flow solutions with smooth solution profiles and treat the discontinuities as continuous solutions with large gradient. The non-dissipative central schemes are preferable for the simulations of smooth broadband turbulence involving small-magnitude features. When simulating flows with sharp discontinuities, the artificial viscosity/diffusivity is usually introduced in the classic central schemes to suppress spurious oscillation across various discontinuities. Nevertheless, the artificial viscosity should be carefully treated in order to avoid overwhelming the small-scale turbulent structures \cite{1,2,3}. Contrary to the central schemes, the upwind schemes consider the flow solutions with discontinuous solution profiles. Since the discontinuous solutions exist at the element interface, the Riemann problems are usually solved to account for the physics of wave propagation. Popular upwind schemes, such as total variation diminishing (TVD) \cite{4}, essentially non-oscillatory (ENO) \cite{5, 6} and weighted ENO (WENO) \cite{7} methods, are able to capture shocks without oscillations. Unfortunately, these traditional upwind schemes are found to be too dissipative to effectively resolve the small flow structures which play essential role in the turbulent interaction process \cite{8}. In previous studies of \cite{9, 10}, the authors demonstrated the importance of reducing numerical dissipation for turbulence simulations by showing that the second-order central scheme performed better than the seventh-order upwind-biased WENO scheme in preserving turbulence energy and spectrum for the explicit LES (ELES) and DNS of compressible turbulence.

Having realized the deficiency of traditional shock-capturing schemes which are based on dissipative upwind schemes, the researchers proposed several methods in aims of simulating compressible turbulence flows.  In the so-called hybrid schemes for compressible turbulence simulations \cite{11,12,13,14,15,16,17}, the large-scale features such as discontinuities are usually captured with nonlinear upwind-biased shock-capturing schemes like TVD and WENO, while the central flux or upwind flux with low-dissipation/dispersion characteristics is introduced to reduce the numerical diffusion on the smooth solution region. Nevertheless, the performance of the hybrid schemes relies on the shock indicator that identifies the discontinuity in the solution. The authors of \cite{18} investigated a wide range of such indicators and concluded that there was no universally better performing method for every problem. Similar with hybrid schemes, several adaptive central-upwind WENO schemes have been designed in \cite{15, LiChen} by recovering the non-dissipative central scheme in smooth region. However, as shown in a very recent work \cite{LiChen}, the adaptive central-upwind schemes may not be robust enough to suppress numerical oscillations across the discontinuity. Besides these central-upwind schemes, extensive works \cite{19,20,21,22,23,24,25,26,27,28} were also done in improving the classical shock-capturing WENO method \cite{7} to minimize the numerical dissipation for bandwidth waves. For instance, early works include optimizing the method of calculating the weights in the WENO scheme \cite{19,20,21} or adding downwind candidate stencil in the reconstruction process \cite{22, 23} to make schemes more centralized. Recent works include monotonicity-preserving (MP) WENO \cite{24, 25}, compact-reconstruction WENO (CRWENO) \cite{26} and targeted ENO (TENO) \cite{27, 28} et al.  It is worth mentioning that many shock-capturing schemes described above have also been extended to other high order discretization framework and implemented as limiting strategies for high speed compressible flow simulations \cite{29,30,31,32}.

%Currently high order shock-capturing schemes implemented in the finite difference (FD) framework may be still the most popular choice for DNS or LES of compressible flows involving strong shocks. 
As indicated in \cite{33}, the numerical dissipation of current shock-capturing schemes mainly comes from the inherent dissipation errors in upwind-biased interpolations, and extra numerical errors introduced by limiting processes which are designed to suppress numerical oscillations. While there have been researches in improving the spectral properties of the shock-capturing scheme by optimizing dispersion and dissipation relation of the scheme \cite{34, 23}, most of the methods are often problem dependent. Also as pointed out in \cite{12}, a small amount of dissipation is needed in order to suppress numerical instabilities caused by under-resolved scale of high wavenumber structures. Therefore, it would be beneficial to design high order numerical schemes with controllable dissipation especially for DNS or LES of compressible turbulence. In recognition of this point of view, the work in \cite{36, 37} developed minimized dispersion and controllable dissipation (MDCD) schemes with the ability of adjusting dissipation without affecting the optimized dispersion properties of the method.

Recently, a novel reconstruction approach named as boundary variation diminishing (BVD) algorithm \cite{38} has been proposed to construct shock capturing schemes in Godunov finite volume method \cite{39,40,41,42}. In the BVD paradigm, the numerical dissipation is effectively reduced by adaptively selecting the reconstruction functions to minimize the variations at the cell boundary. Using proper BVD algorithms and BVD-admissible functions such as high order polynomials and jump-like THINC (Tangent of Hyperbola for INterface Capturing) function, a new class of high order upwind-biased schemes named as PnTm-BVD (polynomial of n-degree and THINC function of m-level reconstruction based on BVD algorithm) has been designed \cite{40,41}.
%and applied to compressible turbulence simulations \cite{33}.
Unlike conventional shock capturing strategies such as TVD and WENO schemes which introduce excessive numerical dissipations through limiting processes and thus undermine the spectral properties of underlying high order interpolations, the PnTm-BVD schemes significantly reduce the numerical errors of limiting processes and are able to retrieve the dispersion and dissipation properties of high-order unlimited polynomials for all wave numbers. With this property, the PnTm-BVD has been applied in compressible turbulence simulations in \cite{33} through an approach known as implicit LES (ILES) \cite{43,44,45} where numerical dissipation is used to model under-resolved flow scale instead of explicit modelling. For instance, the recent work of \cite{42} has shown that the ILES of turbulent flows using higher order PnTm-BVD scheme, as high as 11th-order, is not only more accurate but also more efficient by performing the method on the coarser meshes. However, as an upwind-biased scheme, the inherent and non-tunable numerical dissipation in PnTm-BVD schemes still raises challenging issues to resolve small-scale structures in the DNS of broadband turbulence or to simulate the under-resolved flow with the ILES approach. 

In order to solve multi-scale features in high speed compressible flows, current work has further explored the concept of BVD by introducing the controllable dissipation (CD) algorithm. The BVD with controllable dissipation, denoted by BVD-CD, is then used as the guideline for designing a novel class of multi-scale resolving schemes. The scheme is named as PnTm-BVD-CD (polynomial of n-degree and THINC function of m-level reconstruction based on BVD-CD algorithm) schemes. The PnTm-BVD-CD schemes employ n + 1/n + 2 order upwind/central interpolation based on a centered grid stencil and THINC functions with adaptive steepness as reconstruction candidates. 
%The high order upwind/central interpolation achieves less-dissipative characteristic in the smooth solution region, while the THINC interpolation realizes the non-oscillatory property across various discontinuities.
Through introducing a tunable parameter, the final dissipation property of PnTm-BVD-CD schemes can be effectively controlled between n + 1 order upwind scheme and non-dissipative n + 2 order central scheme. The proposed schemes have at least following desirable characteristics: 1) The proposed schemes can effectively suppress spurious numerical oscillations across the strong shock; 2) The large-scale discontinuous features such as contact, material interface and shear layers can be resolved with less dissipation; 3) The scheme can recover to non-dissipative central interpolation for smooth solution over all wave number, which is important to solve small scale in DNS as well as resolvable scale in ELES; 4) The effectively controllable dissipation algorithm is believed to work for the ILES to solve the under-resolved flow scale. The rest of the paper is organized as follows. Section 2 briefly introduces the governing equation and the FV formulation applied in the work. Section 3 describes the details of BVD-CD algorithms to construct the new dissipation controllable, multi-scale resolving shock capturing schemes. Numerical tests are carried out in Section 4 and some conclusions are drawn in Section 5.

\section{Governing equation and finite volume method}
To simulate high speed compressible viscous flow, the Navier-Stokes equations for a calorically perfect gas are solved:
\begin{eqnarray}
&  & \frac{\partial\rho}{\partial t}+\nabla\cdot\mathbf{m}=0,\label{dens-visd}\\
&  & \frac{\partial\mathbf{m}}{\partial t}+\nabla\cdot\left(\mathbf{m}\otimes\mathbf{u}+p\delta\right)=\nabla \cdot\tau,\label{momt-visd}\\
&  & \frac{\partial E}{\partial t}+\nabla\cdot\left(\mathbf{u}E+p\mathbf{u}\right)=\nabla\cdot(\mathbf{u}\cdot\tau-\mathbf{q}),\label{eneg-visd}
\end{eqnarray}
where $\rho$ is the density, $\mathbf{m}$ is the momentum, $\mathbf{u}$ is the velocity, $\mathbf{p}$ is the pressure, $\delta$ is the unit tensor, $E$ is the total energy, $\tau$ is the viscous stress tensor and $\mathbf{q}$ is the heat flux. The equation of state for ideal gas is applied to closure the above equation system as:
\begin{equation}
 p=(\gamma-1)\rho e   
\end{equation}
in which $\gamma$ is the ratio of specific heats and $e$ is the specific internal energy. With Stokes' hypothesis for a Newtonian
fluid, the viscous stress tensor is calculated as
\begin{equation}
 \tau=2\mu \mathbf{S}-\dfrac{2}{3}\mu (\nabla \cdot \mathbf{u})   
\end{equation}
where strain rate tensor $\mathbf{S}$ is defined as $\mathbf{S}=\dfrac{1}{2}(\nabla \mathbf{u}+(\nabla \mathbf{u})^{T})$ and $\mu$ is the dynamic shear viscosity. Following Fourier's law, the heat flux is defined as $\mathbf{q}=-k\nabla T$ where $k$ is the thermal conductivity and $T$ is the temperature. For simulating the compressible turbulence, the accuracy of discretizing the convective terms has been considered critical \cite{tgref}. Thus, the main challenge is how to design numerical scheme to solve the inviscid Navier-Stokes equations which is also known as Euler equation systems. We use the one-dimensional scalar model equation to illustrate the finite volume method to solve Euler equation systems as
\begin{equation}
\label{eq:scalar}
\frac{\partial q}{\partial t} + \frac{\partial f(q)}{\partial x} = 0,
\end{equation}
where $q$ is the solution variables which can be in characteristic space and $f(q)$ is the flux function.

A standard finite volume method is applied to 1D scalar model equation of Eq.~\ref{eq:scalar} to obtain the numerical solutions. We divide the computational domain into $N$ non-overlapping cell elements, ${\mathcal I}_{i}: x \in [x_{i-1/2},x_{i+1/2} ]$, $i=1,2,\ldots,N$, with a uniform grid spacing $h=\Delta x=x_{i+1/2}-x_{i-1/2}$. the volume-integrated average value $\bar{q}_{i}(t)$ in cell ${\mathcal I}_{i}$ is defined as
\begin{equation}
\bar{q}_{i}(t) 
\approx \frac{1}{\Delta x} \int_{x_{i-1/2}}^{x_{i+1/2}}
q(x,t) \; dx.
\end{equation}
The semi-discrete version of Eq.~(\ref{eq:scalar}) in the finite volume form can be expressed as an ordinary differential equation (ODE)
\begin{equation}
\frac{\partial \bar{q}(t)}{\partial t}  =-\frac{1}{\Delta x}(\tilde{f}_{i+1/2}-\tilde{f}_{i-1/2}),
\end{equation}
where the numerical fluxes $\tilde{f}$ at cell boundaries can be computed by a Riemann solver
\begin{equation}
\tilde{f}_{i+1/2}=f_{i+1/2}^{\text{Riemann}}(q_{i+1/2}^{L},q_{i+1/2}^{R}).
\end{equation}
The left-side value $q_{i+1/2}^{L}$ and right-side value $q_{i+1/2}^{R}$ at cell boundaries are obtained through reconstruction process. The Riemann flux can be generally written in a canonical form as
\begin{equation}
\label{eq:Riemann}
f_{i+1/2}^{\text{Riemann}}(q_{i+1/2}^{L},q_{i+1/2}^{R}) =\frac{1}{2}\left(f(q_{i+1/2}^{L})+f(q_{i+1/2}^{R})\right)-\frac{|a_{i+1/2}|}{2}\left(q_{i+1/2}^{R}-q_{i+1/2}^{L})\right),
\end{equation}
where $a_{i+1/2}$ stands for the characteristic speed of the hyperbolic conservation law. Based on this canonical form, the central schemes lead to non-dissipative flux by considering the flow solutions as smooth solution profiles and regarding discontinuities as a continuous solution with large gradient. Moreover, by controlling the jump $|q_{i+1/2}^{R}-q_{i+1/2}^{L}|$ across the cell boundary, the dissipation added to the flux can be controlled.

\section{Dissipation controllable, multi-scale resolving schemes}
The solution quality is very dependent on how to approximate solution in each cell and reconstruct the $q_{i+1/2}^{L}$ and $q_{i+1/2}^{R}$ which are used to calculate numerical flux. To solve smooth small-scale features, high order polynomials are usually employed to approximate solutions. The chosen polynomials will determine the inherent dissipation and dispersion property of the scheme. In contrary, lower order or monotone functions are more preferable to present discontinuous solutions. Following such idea, upwind-biased high order polynomials and jump-like monotone THINC function are employed as candidate interpolants. The final reconstruction function in each cell is selected from these candidate interpolants with the multi-step BVD algorithm which is in aims of minimizing numerical dissipation. We will review the candidate interpolants before the description of the multi-step BVD algorithm.

\subsection{Candidate interpolant $\mathrm{P}_{n}$: upwind-biased scheme with polynomial of $n$-degree}
A $(n+1)$th order scheme can be constructed from a spatial approximation for the solution in the target cell ${\mathcal I}_{i}$ with a polynomial $\tilde{q}_{i}^{Pn}(x)$ of degree $n$. The $n+1$ unknown coefficients of the polynomial are determined by requiring that $\tilde{q}_{i}^{Pn}(x)$ has the same cell average on each cell over an appropriately selected stencil $S=\{i-n^{-},\dots, i+n^{+}\}$ with $n^{-}+n^{+}=n$ as
\begin{equation}\label{Eq:linearR}
\dfrac{1}{\Delta x}\int_{x_{j-1/2}}^{x_{j+1/2}}\tilde{q}_{i}^{Pn}(x)dx= \bar{q}_{j}, ~~j=i-n^{-},i-n^{-}+1,\dots,i+n^{+}.
\end{equation}
As shown in \cite{7,very3}, $2r-1$ order upwind-biased finite volume schemes can be constructed if the stencil is defined with $n^{-}=n^{+}=r-1$. The unknown coefficients of polynomial of $2r-2$ degree can be then calculated from \eqref{Eq:linearR}. With the polynomial $\tilde{q}_{i}^{Pn}(x)$, high order approximation for reconstructed values at the cell boundaries can be obtained by
\begin{equation}\label{Eq:upwind}
q^{L,Pn}_{i+\frac{1}{2}}=\tilde{q}_{i}^{Pn}(x_{i+\frac{1}{2}}) \ \ {\rm and} \ \  q^{R,Pn}_{i-\frac{1}{2}}=\tilde{q}_{i}^{Pn}(x_{i-\frac{1}{2}}).
\end{equation}
In this work, we apply upwind schemes from fifth order ($r=3$) to ninth order ($r=5$) by using polynomials of 4th, 6th, and 8th degree as one of candidate interpolants. The explicit formulas of $q_{i+1/2}^{L,Pn}$ and $q_{i-1/2}^{R,Pn}$ for (n+1)th-order scheme are detailed follows, 
\begin{itemize}
 \item 5th-order scheme
\begin{equation}\label{Eq:5thuw}
\begin{aligned}
q_{i+1/2}^{L,P4}=\dfrac{1}{30}\bar{q}_{i-2}-\dfrac{13}{60}\bar{q}_{i-1}+\dfrac{47}{60}\bar{q}_{i}+\dfrac{9}{20}\bar{q}_{i+1}-\dfrac{1}{20}\bar{q}_{i+2}, \\
q_{i-1/2}^{R,P4}=\dfrac{1}{30}\bar{q}_{i+2}-\dfrac{13}{60}\bar{q}_{i+1}+\dfrac{47}{60}\bar{q}_{i}+\dfrac{9}{20}\bar{q}_{i-1}-\dfrac{1}{20}\bar{q}_{i-2}.
\end{aligned}
\end{equation}
 \item 7th-order scheme 
\begin{equation}\label{Eq:7thuw}
\begin{aligned}
q_{i+1/2}^{L,P6}=-\dfrac{1}{140}\bar{q}_{i-3}+\dfrac{5}{84}\bar{q}_{i-2}-\dfrac{101}{420}\bar{q}_{i-1}+\dfrac{319}{420}\bar{q}_{i}+\dfrac{107}{210}\bar{q}_{i+1}-\dfrac{19}{210}\bar{q}_{i+2}+\dfrac{1}{105}\bar{q}_{i+3}, \\
q_{i-1/2}^{R,P6}=-\dfrac{1}{140}\bar{q}_{i+3}+\dfrac{5}{84}\bar{q}_{i+2}-\dfrac{101}{420}\bar{q}_{i+1}+\dfrac{319}{420}\bar{q}_{i}+\dfrac{107}{210}\bar{q}_{i-1}-\dfrac{19}{210}\bar{q}_{i-2}+\dfrac{1}{105}\bar{q}_{i-3}.
\end{aligned}
\end{equation}
 \item 9th-order scheme
\begin{equation}\label{Eq:9thuw}
\begin{aligned}
q_{i+1/2}^{L,P8}=&\dfrac{1}{630}\bar{q}_{i-4}-\dfrac{41}{2520}\bar{q}_{i-3}+\dfrac{199}{2520}\bar{q}_{i-2}-\dfrac{641}{2520}\bar{q}_{i-1}+\dfrac{1879}{2520}\bar{q}_{i}\\&+\dfrac{275}{504}\bar{q}_{i+1}-\dfrac{61}{504}\bar{q}_{i+2}+\dfrac{11}{501}\bar{q}_{i+3}-\dfrac{1}{504}\bar{q}_{i+4}, \\
q_{i-1/2}^{R,P8}=&\dfrac{1}{630}\bar{q}_{i+4}-\dfrac{41}{2520}\bar{q}_{i+3}+\dfrac{199}{2520}\bar{q}_{i+2}-\dfrac{641}{2520}\bar{q}_{i+1}+\dfrac{1879}{2520}\bar{q}_{i}\\&+\dfrac{275}{504}\bar{q}_{i-1}-\dfrac{61}{504}\bar{q}_{i-2}+\dfrac{11}{501}\bar{q}_{i-3}-\dfrac{1}{504}\bar{q}_{i-4}.
\end{aligned}
\end{equation}
\end{itemize}

\subsubsection{Candidate interpolant $\mathrm{T}_{m}$: THINC function with $m$-level steepness}
To present the solution in discontinuous regions, we introduce another candidate interpolation function is the THINC interpolation \cite{xiao_thinc,xiao_thinc2}.  
The piecewise THINC reconstruction function is written as
\begin{equation} \label{eq:THINC}
\tilde{q}_{i}^{T}(x)=\bar{q}_{min}+\dfrac{\bar{q}_{max}}{2} \left[1+\theta~\tanh \left(\beta \left(\dfrac{x-x_{i-1/2}}{x_{i+1/2}-x_{i-1/2}}-\tilde{x}_{i}\right)\right)\right],
\end{equation} 
where $\bar{q}_{min}=\min(\bar{q}_{i-1},\bar{q}_{i+1})$, $\bar{q}_{max}=\max(\bar{q}_{i-1},\bar{q}_{i+1})-\bar{q}_{min}$ and 
$\theta=sgn(\bar{q}_{i+1}-\bar{q}_{i-1})$.The unknown $\tilde{x}_{i}$ is computed from constraint condition $\displaystyle \bar{q}_{i} = \frac{1}{\Delta x} \int_{x_{i-1/2}}^{x_{i+1/2}} \tilde{q}^{T}_{i}(x) dx$. The jump thickness is controlled by the parameter $\beta$ which determines the steepness. Given the reconstruction function $\tilde{q}_{i}^{T}(x)$, we calculate the boundary values $q_{i+1/2}^{L,T}$ and $q_{i-1/2}^{R,T}$ by $q_{i+1/2}^{L,T}=\tilde{q}_{i}^{T}(x_{i+1/2})$ and $q_{i-1/2}^{R,T}=\tilde{q}_{i}^{T}(x_{i-1/2})$ respectively.

To present discontinuities with different steepness, we use THINC functions with $\beta$ of $m$-level. Unlike the high order polynomials, the THINC function can realize non-oscillatory as well as less-dissipative reconstructions for discontinuities.  A THINC reconstruction function $\tilde{q}_{i}^{Tk}(x)$ with $\beta_{k}$ gives the reconstructed values $q_{i+1/2}^{L,Tk}$ and $q_{i-1/2}^{R,Tk}$, ($k=1,2,\dots,m$). We will use $m$ up to three in present study.

\subsection{The BVD-CD algorithm} 
In the $\mathrm{P}_{n}\mathrm{T}_{m}-\mathrm{BVD}-\mathrm{CD}$ schemes, reconstruction values are determined from the candidate interpolants with $k$-stage $(k=m+1)$ BVD-CD algorithm so as to minimize the jumps at cell boundaries as small as possible. We denote the reconstruction values at the cell boundary $i+1/2$ after the $k$-th stage BVD-CD as $q_{i+1/2}^{L,<k>}$ and $q_{i+1/2}^{R,<k>}$.
%We denote the reconstruction function in the target cell ${\mathcal I}_{i}$ after the $k$-th stage BVD as $\tilde{q}_{i}^{<k>}(x)$.  

The $k$-stage $(k=m+1)$ BVD-CD algorithm is formulated as follows. 

\begin{description} 
	\item{\bf (I):  Initial stage $(k=0)$:}  
	\begin{description} 
		\item (I-I) As the first step, use the high-order upwind scheme as the base reconstruction scheme and initialize the reconstructed function as  $\tilde{q}_{i}^{<0>}(x)=\tilde{q}_{i}^{Pn(x)}$. 
		%As the first step, use the linear high-order upwind scheme as the base reconstruction scheme and initialize the reconstructed values for all cell as  $q_{i+1/2}^{L,<0>}=\tilde{q}_{i}^{Pn}(x_{i+\frac{1}{2}})$ and $q_{i+1/2}^{R,<0>}=\tilde{q}_{i+1}^{Pn}(x_{i+\frac{1}{2}})$.  
	\end{description} 
\end{description} 

\begin{description} 
	\item{\bf (II) Limiting process at the intermediate BVD-CD stage $(k=1,\dots,m-1)$: }  
	\begin{description} 
	%	\item (II-I) Set $\tilde{q}_{i}^{<k>}(x)=\tilde{q}_{i}^{<k-1>}(x)$
	%	\item (II-I) Calculate the TBV values for target cell ${\mathcal I}_{i}$ from the previous stage $k-1$
	%	\begin{equation}\label{Eq:TBVp4}
	%	TBV_{i}^{<k-1>}=\big|q_{i-1/2}^{L,<k-1>}-q_{i-1/2}^{R,<k-1>}\big|+\big|q_{i+1/2}^{L,<k-1>}-q_{i+1/2}^{R,<k-1>} \big|
	%	\end{equation} 
	
\item (II-I) Set $\tilde{q}_{i}^{<k>}(x)=\tilde{q}_{i}^{<k-1>}(x)$
		\item (II-II) Calculate the TBV values for target cell ${\mathcal I}_{i}$ from the reconstruction of $\tilde{q}_{i}^{<k>}(x)$
		\begin{equation}\label{Eq:TBVp4}
		TBV_{i}^{<k>}=\big|q_{i-1/2}^{L,<k>}-q_{i-1/2}^{R,<k>}\big|+\big|q_{i+1/2}^{L,<k>}-q_{i+1/2}^{R,<k>} \big|
		\end{equation} 
		and from the THINC function $\tilde{q}_{i}^{Tk}(x)$ with a  steepness $\beta_{k}$ as    
		\begin{equation}\label{Eq:TBVTs}
		TBV_{i}^{Tk}=\big|q_{i-1/2}^{L,Tk}-q_{i-1/2}^{R,Tk}\big|+\big|q_{i+1/2}^{L,Tk}-q_{i+1/2}^{R,Tk} \big|.
		\end{equation} 
		\item (II-III) Modify the reconstruction function for cells $i-1$, $i$ and $i+1$ according to the following BVD algorithm
		\begin{equation}\label{Eq:BVDlim-1}
		\tilde{q}_{j}^{<k>}(x)=\tilde{q}_{j}^{Tk}(x), \ j=i-1,i,i+1;~~~{\rm if } \ \ TBV_{i}^{Tk}  < TBV_{i}^{<k>}.
		\end{equation} 
		\item (II-IV) Compute the reconstructed values on the left-side of $x_{i+\frac{1}{2}}$ and the right-side of $x_{i-\frac{1}{2}}$  respectively by 
		\begin{equation}
		q^{L,<k>}_{i+\frac{1}{2}}=\tilde{q}_{i}^{<k>}(x_{i+\frac{1}{2}}) \ \ {\rm and} \ \  q^{R,<k>}_{i-\frac{1}{2}}=\tilde{q}_{i}^{<k>}(x_{i-\frac{1}{2}}).
		\end{equation}

	\end{description} 
\end{description} 

\begin{description} 
	\item {\bf  (III) The dissipation control process at BVD-CD stage $(k=m)$:}
	
    Given the reconstruction values from the above stage $m-1$, for each cell boundary $i+\frac{1}{2}$ modify the reconstruction values as 
    		\begin{equation}
	\left\{
		\begin{array}{l}
		q^{L,<k>}_{i+\frac{1}{2}}=\lambda q^{L,Pn}_{i+\frac{1}{2}}+(1.0-\lambda)q^{R,Pn}_{i+\frac{1}{2}}; \\
		q^{R,<k>}_{i+\frac{1}{2}}=\lambda q^{R,Pn}_{i+\frac{1}{2}}+(1.0-\lambda)q^{L,Pn}_{i+\frac{1}{2}};
		\end{array}
		\right. ~~~~~{\rm if }~~ q^{L,<m-1>}_{i+\frac{1}{2}}=q^{L,Pn}_{i+\frac{1}{2}}~~{\rm and }~~q^{R,<m-1>}_{i+\frac{1}{2}}=q^{R,Pn}_{i+\frac{1}{2}} 
		\end{equation}
    
\end{description}

\begin{description} 
	\item {\bf  (IV) The large scale discontinuity sharpening process at the final BVD-CD stage $(k=m+1)$:} 
	\begin{description} 
		\item (III-I) 
		Given the reconstruction values from the above stage $m$, compute the TBV using the reconstructed cell boundary values from previous stage by     
		\begin{equation}\label{Eq:TBVlim}
		TBV_{i}^{<m>}=\big|q_{i-1/2}^{L,<m>}-q_{i-1/2}^{R,<m>}\big|+\big|q_{i+1/2}^{L,<m>}-q_{i+1/2}^{R,<m>} \big|, 
		\end{equation}
		and the TBV for THINC function of $\beta_{m}$ by 
		\begin{equation}\label{Eq:TBVtl}
		TBV_{i}^{Tm}=\big|q_{i-1/2}^{L,Tm}-q_{i-1/2}^{R,Tm}\big|+\big|q_{i+1/2}^{L,Tm}-q_{i+1/2}^{R,Tm} \big|. 
		\end{equation}
		\item (III-II)  Determine the final reconstruction function for cell ${\mathcal I}_{i}$ using the  BVD algorithm as
		%\begin{equation}\label{Eq:BVDlim-2}
		%\tilde{q}_{i}^{<\text{II}>}(x)=\tilde{q}_{j}^{Ts}(x), \ j=i-1,i,i+1~~~{\rm if } \ \ TBV_{i}^{Ts}  < TBV_{i}^{p4}.
		%\end{equation} 
		\begin{equation}
		\tilde{q}_{i}^{<m+1>}(x)=\left\{
		\begin{array}{l}
		\tilde{q}_{i}^{Tm};~~~{\rm if } \ \ TBV_{i}^{Tm}  < TBV_{i}^{<m>}, \\
		\tilde{q}_{i}^{<m>};~~~~\mathrm{otherwise}
		\end{array}
		\right..
		\end{equation}
		
		\item (III-III) Compute the reconstructed values on the left-side of $x_{i+\frac{1}{2}}$ and the right-side of $x_{i-\frac{1}{2}}$  respectively by 
		\begin{equation}
		q^{L}_{i+\frac{1}{2}}=\tilde{q}_{i}^{<m+1>}(x_{i+\frac{1}{2}}) \ \ {\rm and} \ \  q^{R}_{i-\frac{1}{2}}=\tilde{q}_{i}^{<m+1>}(x_{i-\frac{1}{2}}).
		\end{equation}

	\end{description} 
\end{description} 

\begin{description}
\item {Remark 1. } 
With $\lambda$ decreased from 1.0 to 0.5, the interpolation becomes more centralized. With this parameter, the dissipation property can be effectively controlled between high order upwind scheme and non-dissipative central scheme. 
\item {Remark 2. } 
The controllable dissipation is believed to work for solving under-resolved small-scale turbulence in ILES, where controllable numerical dissipation can be used in replace of the explicit sub-grid scale (SGS) model. With $\lambda=0.5$, $n+2$ order non-dissipative central interpolation can be recovered, which is believed to work for solving resolvable small-scale in DNS or ELES.
\item {Remark 3. } 
As shown in numerical tests, the BVD-CD algorithm can solve large-scale discontinuities with non-essential oscillation property and reproduce sharp discontinuous solution.
%\item {Remark 2. }   
%The sequence of the dissipation control process at the stage $m$ and the discontinuity sharping process at the stage $m+1$ can be swapped if the discontinuity sharpening is the priority.
 
\end{description}

In this study we first propose schemes with $\lambda=0.5$ and test sixth order scheme with $\mathrm{P}_{4}\mathrm{T}_{2}-\mathrm{BVD-CD}$, eighth order scheme with $\mathrm{P}_{6}\mathrm{T}_{3}-\mathrm{BVD-CD}$ and tenth order scheme with $\mathrm{P}_{8}\mathrm{T}_{3}-\mathrm{BVD-CD}$. According to previous study in \cite{41}, in all tests of the present study we use $\beta_{1}=1.1$ and $\beta_{2}=1.6$ for $\mathrm{P}_{4}\mathrm{T}_{2}-\mathrm{BVD-CD}$, and  $\beta_{1}=1.2$, $\beta_{2}=1.1$ and $\beta_{3}=1.6$ for $\mathrm{P}_{n}\mathrm{T}_{3}-\mathrm{BVD-CD}~(n=6,8)$ schemes. As well known, the high speed compressible flow is challenging for very low-dissipation schemes. Thus we mainly test the central scheme with $\lambda=0.5$ to show the performance of the proposed scheme. The proposed scheme will also be tested with ILES approach. 

\section{Numerical experiments}
In this section, we verify the performance of proposed $\mathrm{P}_{n}\mathrm{T}_{m}-\mathrm{BVD-CD}$ schemes in simulating in both Euler and NS equations. We choose an extreme case with $\lambda=0.5$ which recovers to central scheme for smooth region.  We will test the scheme with $\lambda=0.5$ through non-broadband 1D and 2D Euler equations containing discontinuities. Then the simulation results of NS equations including broadband compressible turbulence problems will be given through ILES approach. Numerical results are compared with high order mapped WENO (WENOM) \cite{WENOM} schemes or a very recent upwind-central WENO scheme \cite{LiChen}. For the time scheme, $n+1$ order linear strong-stability-preserving Runge-Kutta algorithms developed in \cite{time} are used. The CFL number $0.4$ is used in our tests unless specifically noted.

\subsection{Spectral property}
\begin{figure} [h]
	\begin{center}
    \subfigure[6th order scheme]
	{\centering\includegraphics[width=.32\textwidth,trim={0.5cm 0.5cm 0.5cm 0.5cm},clip]{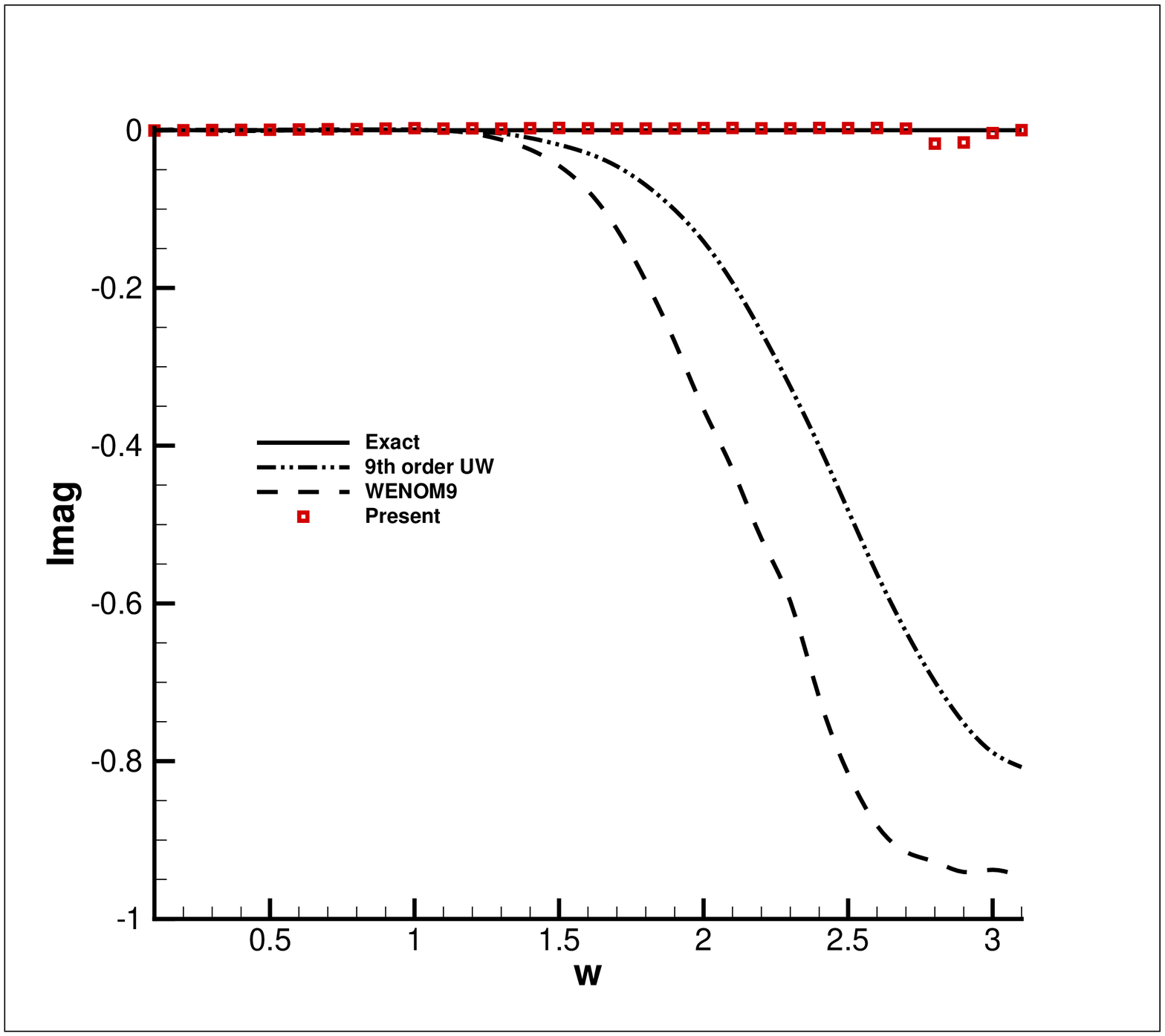}}
    \subfigure[8th order scheme]
	{\centering\includegraphics[width=.32\textwidth,trim={0.5cm 0.5cm 0.5cm 0.5cm},clip]{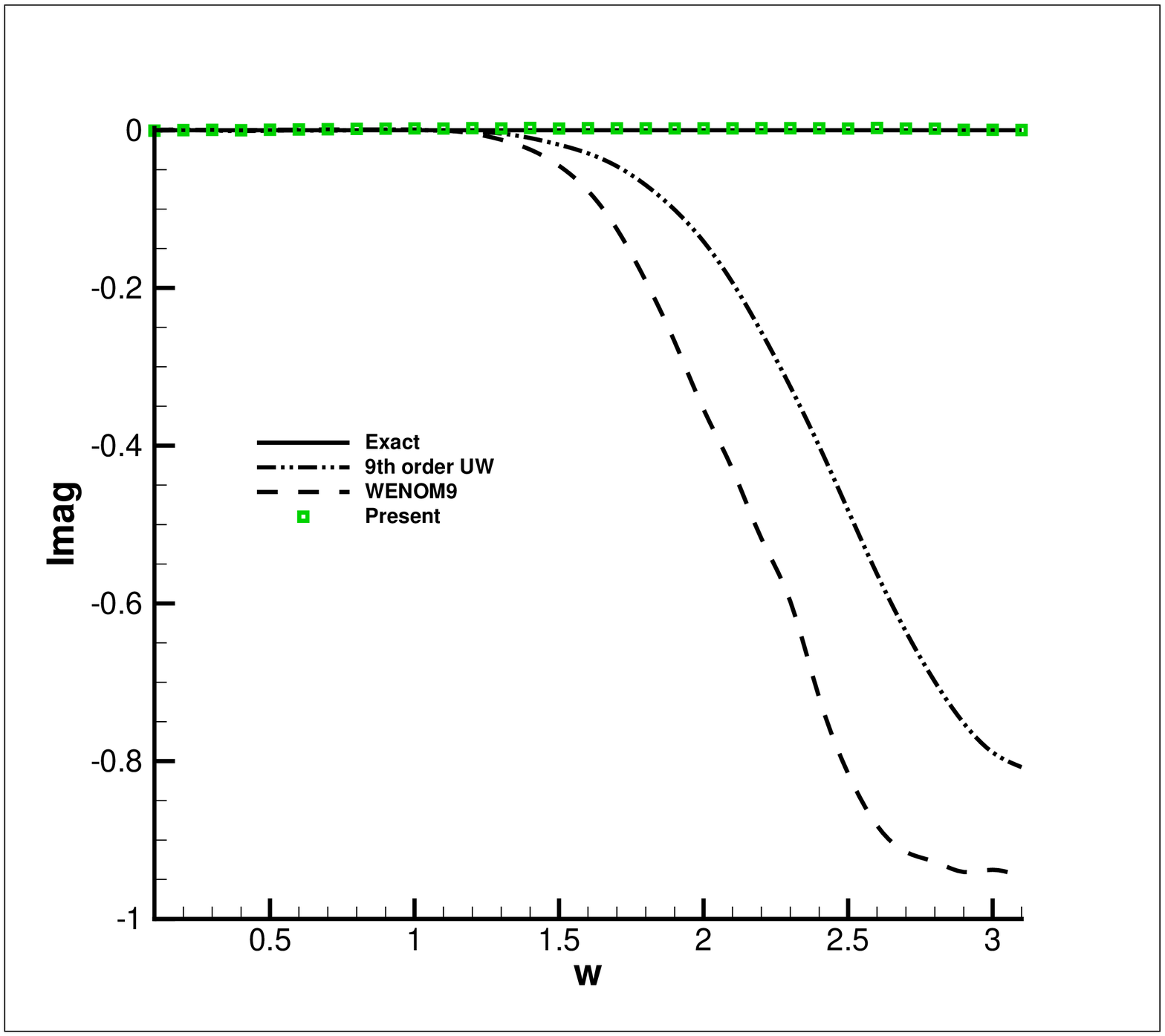}}
	\subfigure[10th order scheme]
	{\centering\includegraphics[width=.32\textwidth,trim={0.5cm 0.5cm 0.5cm 0.5cm},clip]{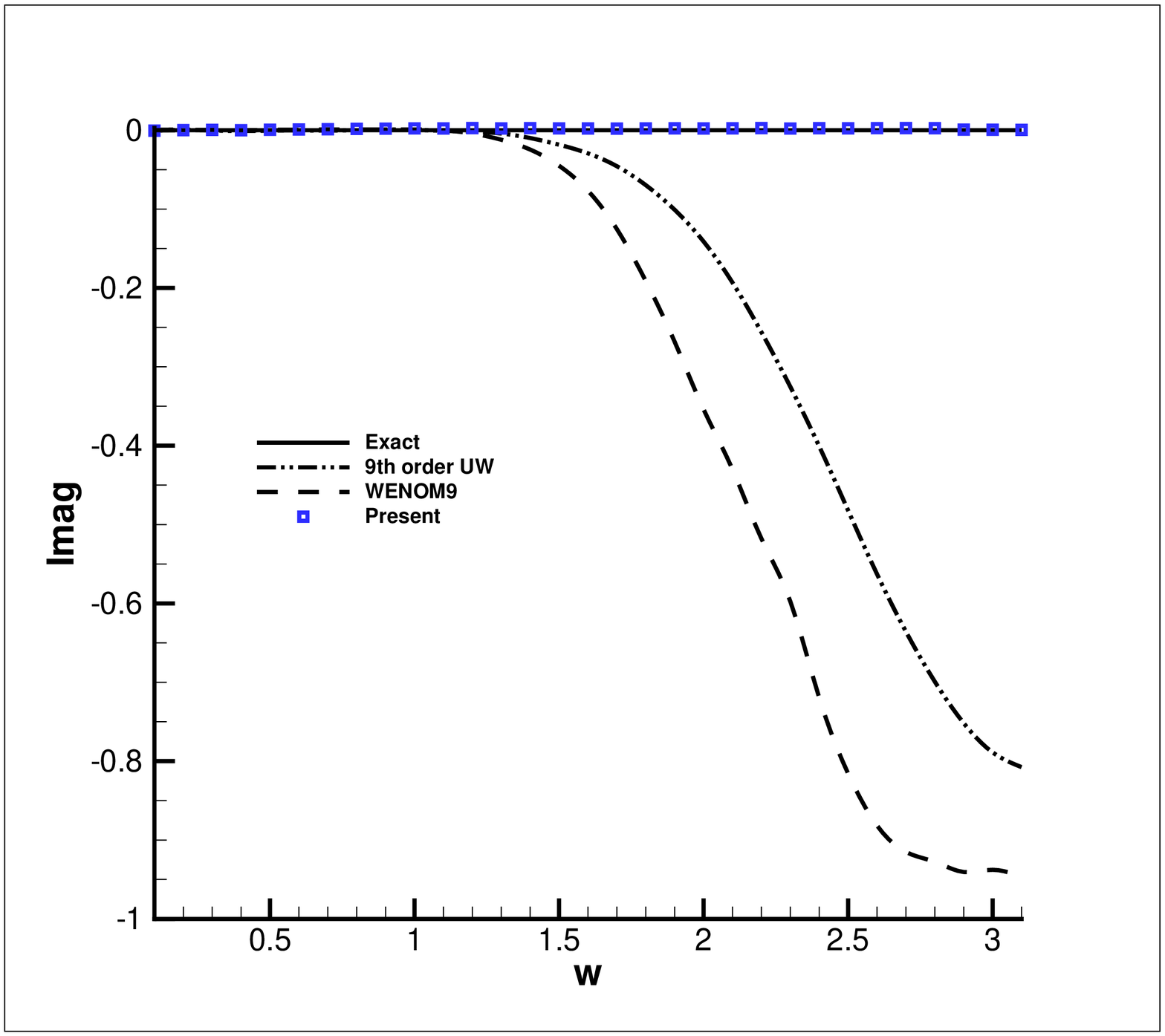}}
		\protect\caption{Approximate dissipation properties for different schemes are analyzed by imaginary parts of modified wavenumber. Comparisons among linear upwind schemes, WENOM and proposed $\mathrm{P}_{n}\mathrm{T}_{m}-\mathrm{BVD-CD}$ schemes.
			\label{fig:ADRdissipation}}
	\end{center}	
\end{figure}

The approximate dispersion relation (ADR) method described in \cite{adr} is applied to study the spectral property of the proposed scheme.
The numerical dissipation property of the ninth order linear upwind schemes, WENOM and proposed $\mathrm{P}_{n}\mathrm{T}_{m}-\mathrm{BVD-CD}$ schemes are shown in Fig.~\ref{fig:ADRdissipation}. It is obvious that there is inherent numerical dissipation in upwind-biased schemes despite of very high order. Moreover,the spectral property of ninth order WENOM schemes is inferior to the same order linear scheme for high wavenumber regime. These discrepancies are partly caused by WENO non-linear adaptation which may regard high wavenumber waves as discontinuities. It is noteworthy that despite of having developed more accurate WENO schemes in recent year, to recover the upwind-biased scheme in whole wavenumber regime hasn't been fully solved through WENO methodology. 
On the contrary, through newly devised BVD-CD processes the proposed schemes almost realize the non-dissipative property for all wavenumbers. Although it can be observed that there is still very small numerical dissipation in high wavenumber region for the sixth order scheme, the proposed eighth order and tenth order schemes retrieve corresponding non-dissipative high order central schemes. In following sections, we will show that the proposed schemes are also accurate and robust for shock capturing even with such non-dissipative central scheme for smooth regions. 

\begin{figure} [h]
	\begin{center}
    \subfigure[6th order scheme]
	{\centering\includegraphics[width=.32\textwidth,trim={0.5cm 0.5cm 0.5cm 0.5cm},clip]{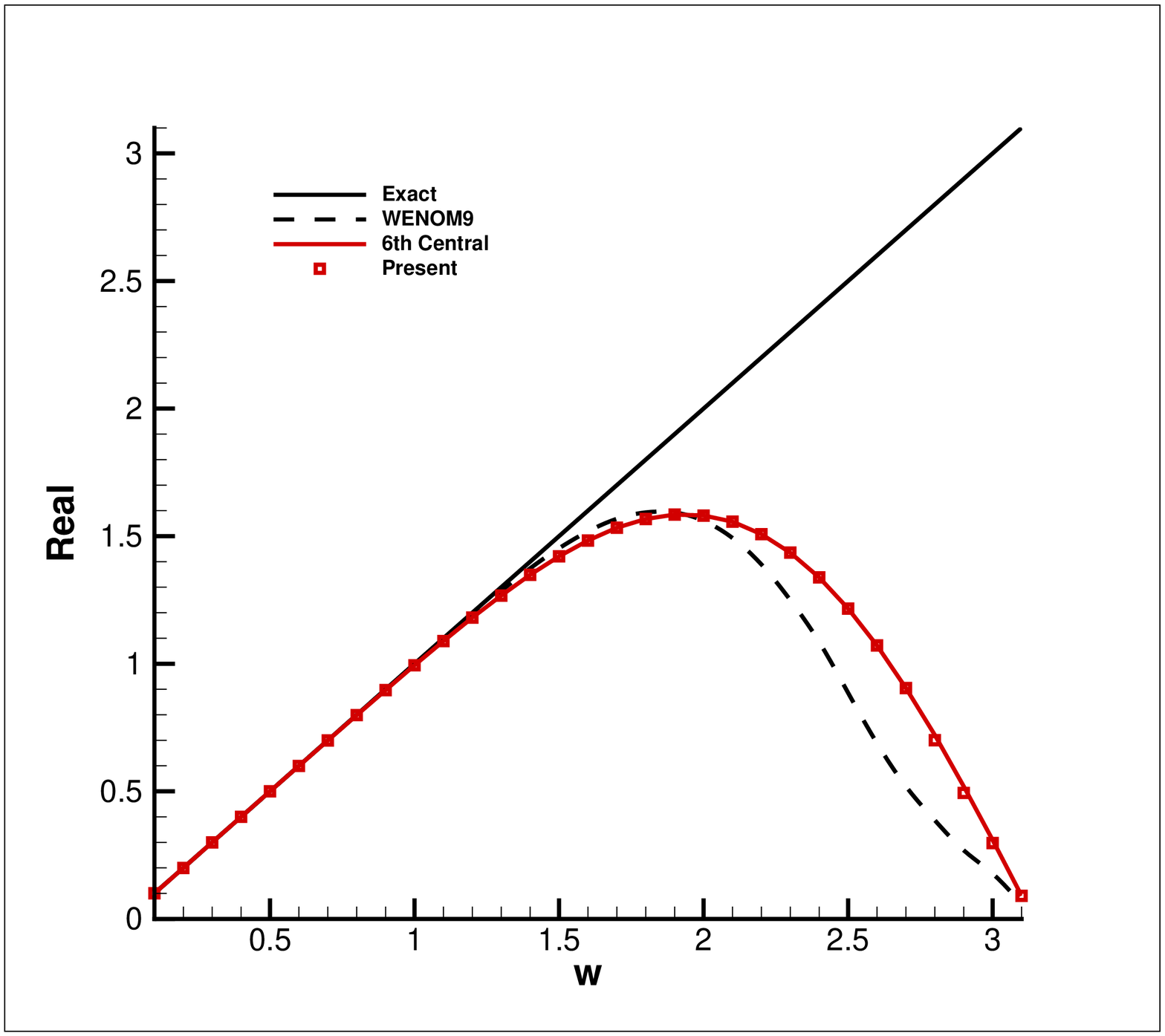}}
    \subfigure[8th order scheme]
	{\centering\includegraphics[width=.32\textwidth,trim={0.5cm 0.5cm 0.5cm 0.5cm},clip]{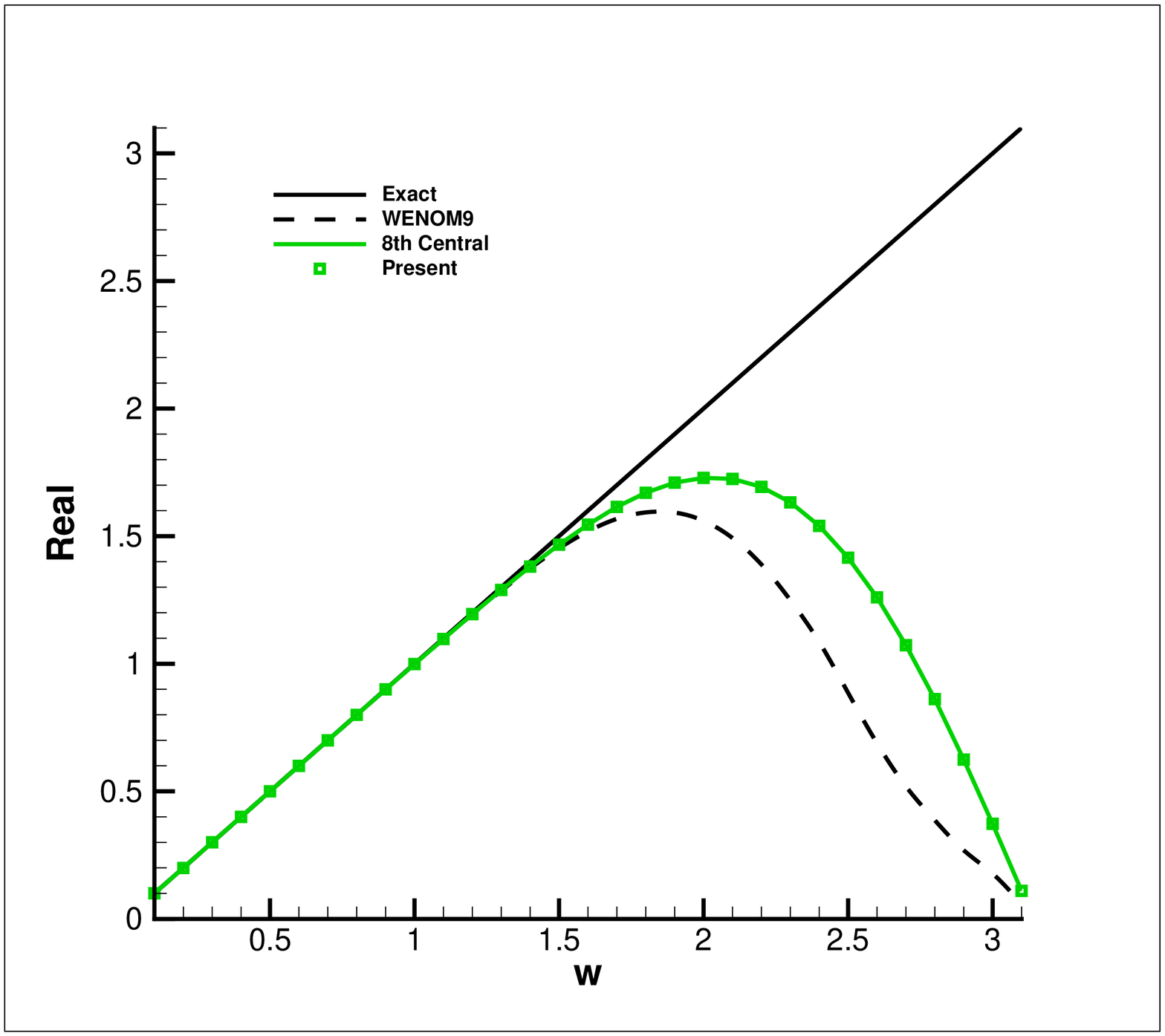}}
	\subfigure[10th order scheme]
	{\centering\includegraphics[width=.32\textwidth,trim={0.5cm 0.5cm 0.5cm 0.5cm},clip]{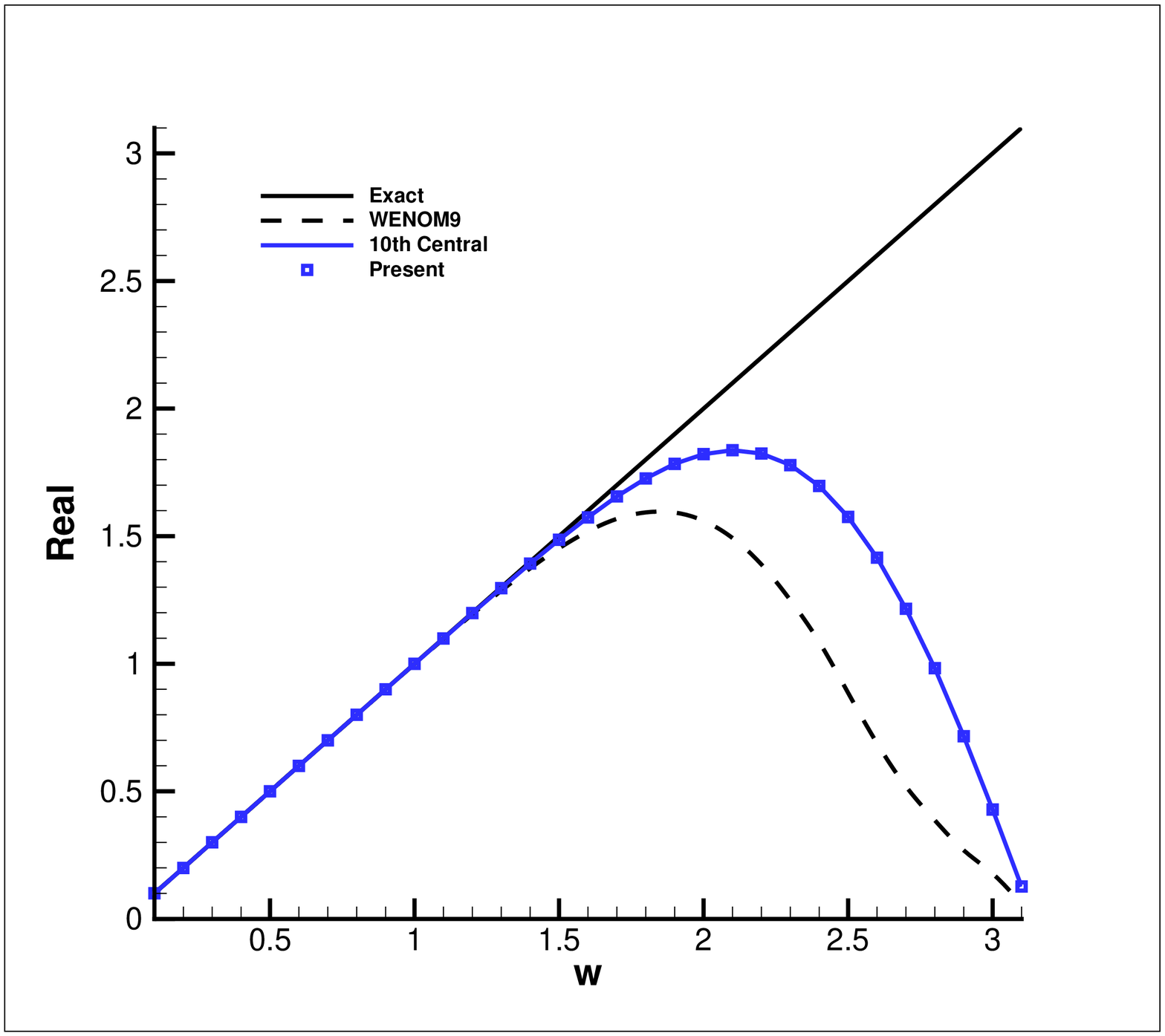}}
		\protect\caption{Approximate dispersion property for different schemes are analyzed by real parts of modified wavenumber. Comparisons among central schemes, WENOM and $\mathrm{P}_{n}\mathrm{T}_{m}-\mathrm{BVD-CD}$ schemes.
			\label{fig:ADRdispersion}}
	\end{center}	
\end{figure}

In Fig.~\ref{fig:ADRdispersion}, the dispersion property of proposed schemes are presented by plotting the real parts of modified wavenumbers. All proposed schemes show the almost same dispersion property as their corresponding high order central schemes. Moreover, the dispersion properties of proposed schemes surpass those of very high order WENOM schemes especially in high wavenumber regions. 

In a summary, the proposed schemes are almost able to retrieve the non-dissipative central schemes for all resolvable wavenumbers in smooth region, which are important for simulating turbulence involving broadband flow structures through DNS or ELES. It can be expected that by adjusting parameter $\lambda$, dissipation property of proposed schemes can be exactly controlled between central scheme and their underlying upwind schemes. It is also noted that the dispersion property has been improved as order increased. To further improve the dispersion property which is also important for turbulence flow, optimization processes of dissipation and dispersion \cite{27,28} or compact schemes \cite{compact1,compact2} will be considered to construct low-dispersion schemes in our future work.  

\subsection{Accuracy test for advection of one-dimensional sine wave}\label{accuracy1}
The convergence rate of the proposed $\mathrm{P}_{n}\mathrm{T}_{m}-\mathrm{BVD-CD}$ schemes is studies with an advection test of smooth profile on gradually refined grids. The initial smooth distribution was given by
\begin{equation}
q\left(x\right)=\sin\left(2\pi x\right), \ x\in\left[-1,1\right].
\end{equation}
We ran the computation for one period (at $t=2.0$) and summarized the numerical errors and the convergence rates for $\mathrm{P}_{n}\mathrm{T}_{m}-\mathrm{BVD-CD}$ schemes in Table \ref{Tab:rate}. As expected, the proposed $\mathrm{P}_{n}\mathrm{T}_{m}-\mathrm{BVD-CD}$ schemes achieved $n+2$th order convergence rates when grid elements were gradually refined. Compared with our previous work \cite{41}, the $L_{1}$ and $L_{\infty}$ errors have been significantly reduced through newly designed dissipation control processes and the convergence rate has been improved to one order higher. It can also be verified easily that the $L_{1}$ and $L_{\infty}$ errors from the proposed schemes are exactly same as those calculated by $n+2$th order central schemes, which is in line with the conclusion from the spectral property analysis in the previous section. 

\begin{table}[]
	\centering
	\caption{Numerical errors and convergence rate for linear advection test. Results are computed by the proposed $\mathrm{P}_{n}\mathrm{T}_{m}-\mathrm{BVD-CD}$ schemes.}
	\label{Tab:rate}
	\begin{tabular}{l|lllll}
		\hline
		Schemes                                      & Mesh & $L_{1}$ errors & $L_{1}$ order & $L_{\infty}$ errors & $L_{\infty}$ order \\ \hline
				\multirow{5}{*}{$\mathrm{P_{4}T_{2}}$-BVD-CD}
& 20   & $3.099\times10^{-3}$ &  & $4.788\times10^{-3}$ & \\		                   
& 40   & $5.389\times10^{-5}$ & 5.85 & $8.327\times10^{-5}$ & 5.85\\
& 80   & $8.543\times10^{-7}$ & 5.98 & $1.336\times10^{-6}$ & 5.96 \\
& 160  & $1.340\times10^{-8}$ & 5.99 & $2.102\times10^{-8}$ & 5.99  \\ \hline
				\multirow{5}{*}{$\mathrm{P_{6}T_{3}}$-BVD-CD}
& 20   & $2.634\times10^{-4}$ &  & $4.070\times10^{-4}$ & \\		                   
& 40   & $1.173\times10^{-6}$ & 7.81 & $1.812\times10^{-6}$ & 7.81\\
& 80   & $4.675\times10^{-9}$ & 7.97 & $7.313\times10^{-9}$ & 7.95 \\
& 160  & $1.835\times10^{-11}$ &7.99 & $2.880\times10^{-11}$ & 7.99 \\ \hline
				\multirow{5}{*}{$\mathrm{P_{8}T_{3}}$-BVD-CD}
& 20   & $2.290\times10^{-5}$ &  & $3.538\times10^{-5}$ & \\		                     
& 40   & $2.610\times10^{-8}$ & 9.77 & $4.033\times10^{-8}$ & 9.78\\
& 80   & $2.616\times10^{-11}$ & 9.96 & $4.093\times10^{-11}$ & 9.94 \\
& 160  & $2.547\times10^{-14}$ & 10.00 & $4.374\times10^{-14}$ & 9.87 \\ \hline
			
	\end{tabular}
\end{table}

\subsection{Accuracy test for advection of a smooth profile containing critical points}\label{accuracy2}
Initial distributions containing critical points are more challenging for numerical schemes to distinguish smooth and truly discontinuous profiles. For example, at critical points where the high order derivatives do not simultaneously vanish, WENO-type schemes may not reach their formal order of accuracy. In this test, the initial condition is given following the work \cite{WENOM} by 
\begin{equation}
q\left(x\right)=\sin(\pi x-\dfrac{\sin(\pi x)}{\pi}), \ x\in\left[-1,1\right].
\end{equation}
The computation was conducted for four periods ($t=8$). We summarize the numerical errors $L_{1}$ and $L_{\infty}$ of the proposed scheme in Table \ref{Tab:critical}. As shown in results, the proposed algorithm keeps the expected convergence rate even for solution containing critical points. This test verifies again the smooth and discontinuous solutions can be effectively distinguished by the BVD-CD algorithm.

%%%critical point
\begin{table}[]
	\centering
	\caption{Numerical errors and convergence rate for advection of a smooth profile containing critical points. Results are calculated by the proposed schemes.}
	\label{Tab:critical}
	\begin{tabular}{l|lllll}
		\hline
		Schemes                                      & Mesh & $L_{1}$ errors & $L_{1}$ order & $L_{\infty}$ errors & $L_{\infty}$ order \\ \hline
				\multirow{5}{*}{$\mathrm{P_{4}T_{2}}$-BVD-CD}
& 20   & $3.309\times10^{-3}$ &  & $7.759\times10^{-3}$ & \\		                   
& 40   & $4.805\times10^{-5}$ & 6.11 & $1.542\times10^{-4}$ & 5.65\\
& 80   & $7.823\times10^{-7}$ & 5.94 & $2.554\times10^{-6}$ & 5.92 \\
& 160  & $1.230\times10^{-8}$ & 5.99 & $4.052\times10^{-8}$ & 5.98  \\ \hline
				\multirow{5}{*}{$\mathrm{P_{6}T_{3}}$-BVD-CD}
& 20   & $4.528\times10^{-4}$ &  & $1.299\times10^{-3}$ & \\		                   
& 40   & $2.242\times10^{-6}$ & 7.66 & $7.337\times10^{-6}$ & 7.47\\
& 80   & $9.401\times10^{-9}$ & 7.90 & $3.172\times10^{-8}$ & 7.85 \\
& 160  & $3.935\times10^{-11}$ &7.90 & $1.303\times10^{-10}$ & 7.93 \\ \hline
				\multirow{5}{*}{$\mathrm{P_{8}T_{3}}$-BVD-CD}
& 20   & $9.775\times10^{-5}$ &  & $2.888\times10^{-4}$ & \\		                     
& 40   & $1.389\times10^{-7}$ & 9.46 & $4.748\times10^{-7}$ & 9.25\\
& 80   & $1.450\times10^{-10}$ & 9.90 & $5.425\times10^{-10}$ & 9.77 \\
& 160  & $1.710\times10^{-13}$ & 9.73 & $6.899\times10^{-13}$ & 9.62 \\ \hline
	\end{tabular}
\end{table}

\subsection{Advection of complex waves}
\begin{figure}
	\begin{center}
    \subfigure[6th order scheme]
	{\centering\includegraphics[width=.32\textwidth,trim={0.5cm 0.5cm 0.5cm 0.5cm},clip]{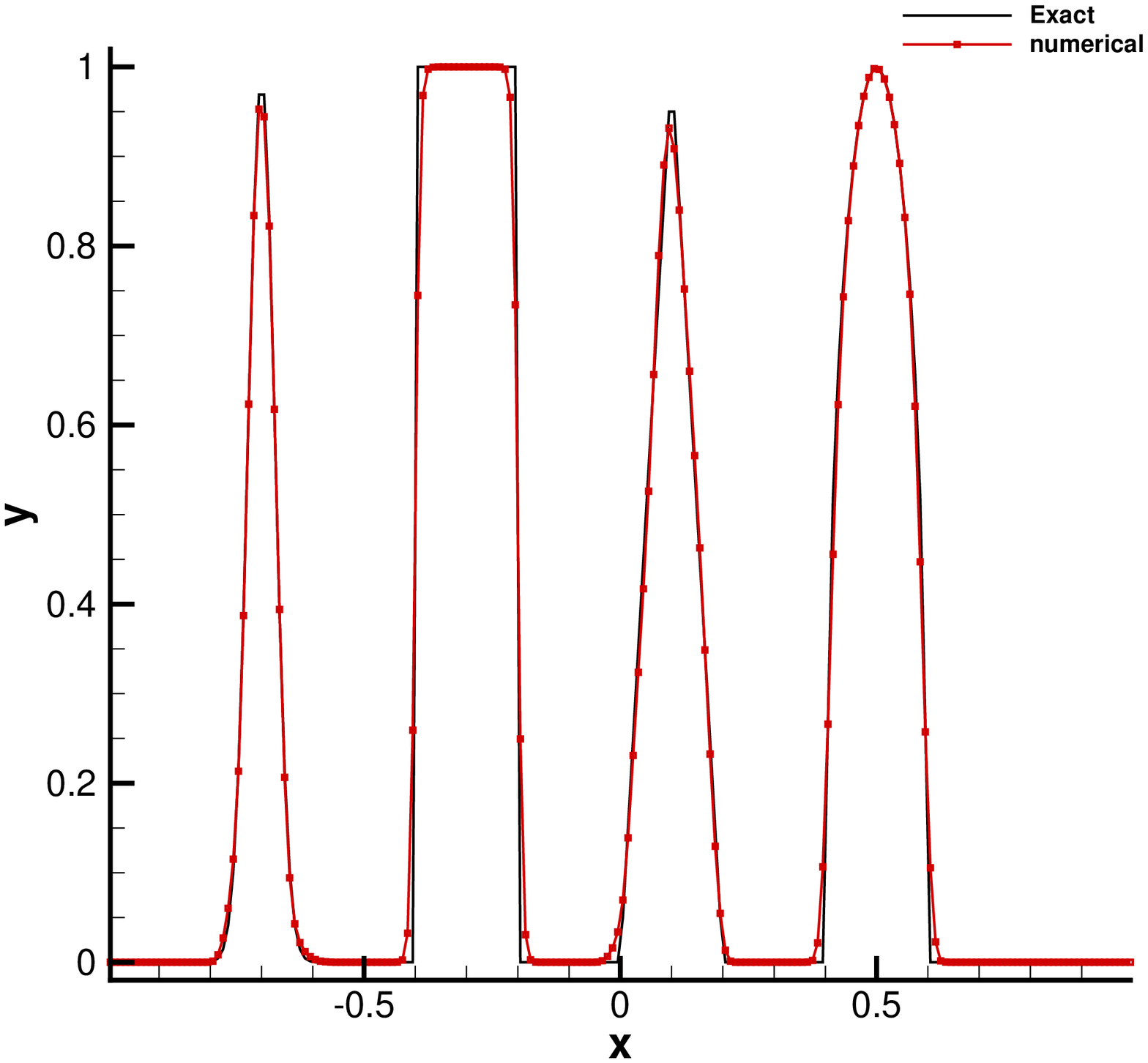}}
    \subfigure[8th order scheme]
	{\centering\includegraphics[width=.32\textwidth,trim={0.5cm 0.5cm 0.5cm 0.5cm},clip]{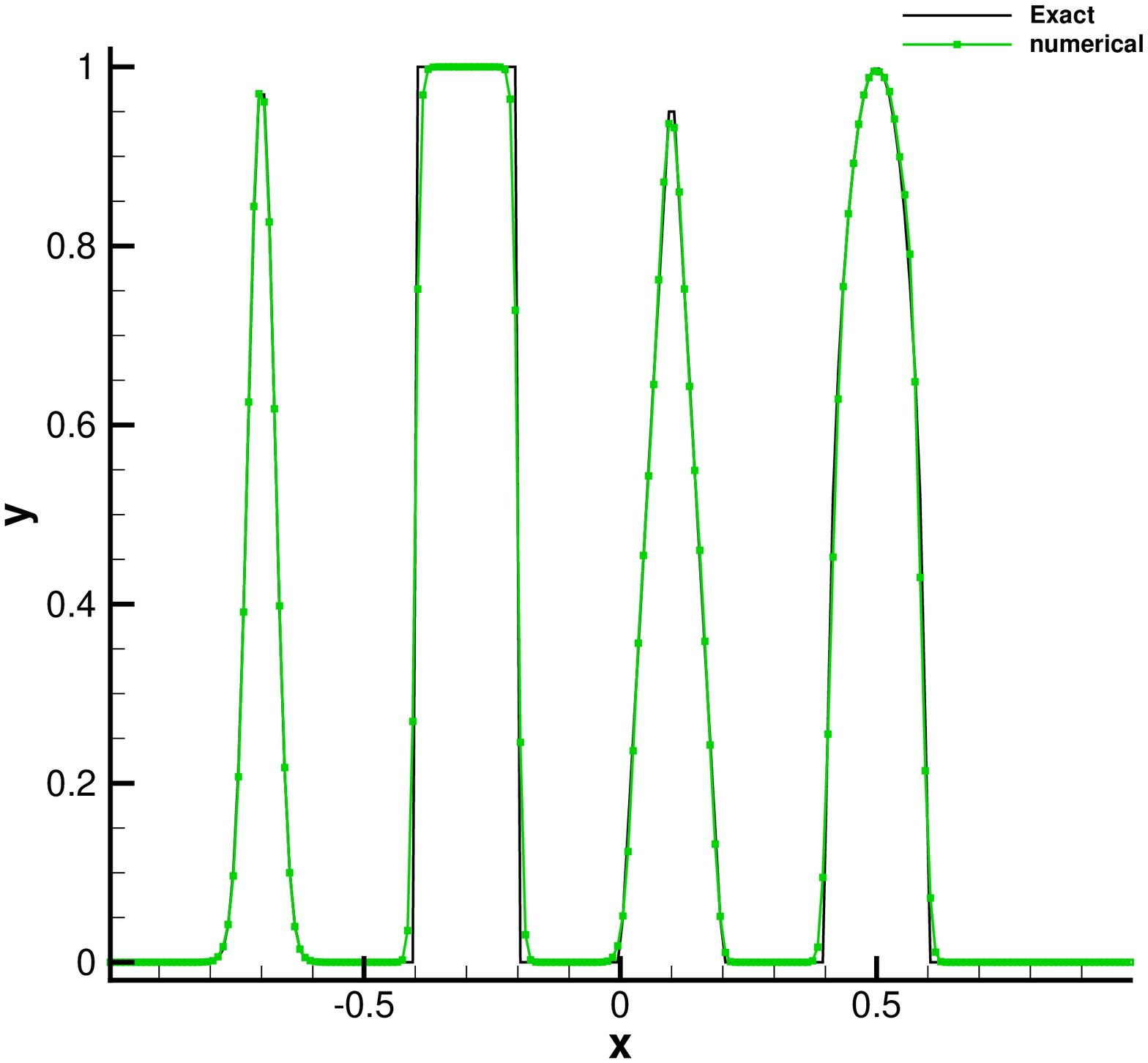}}
	\subfigure[10th order scheme]
	{\centering\includegraphics[width=.32\textwidth,trim={0.5cm 0.5cm 0.5cm 0.5cm},clip]{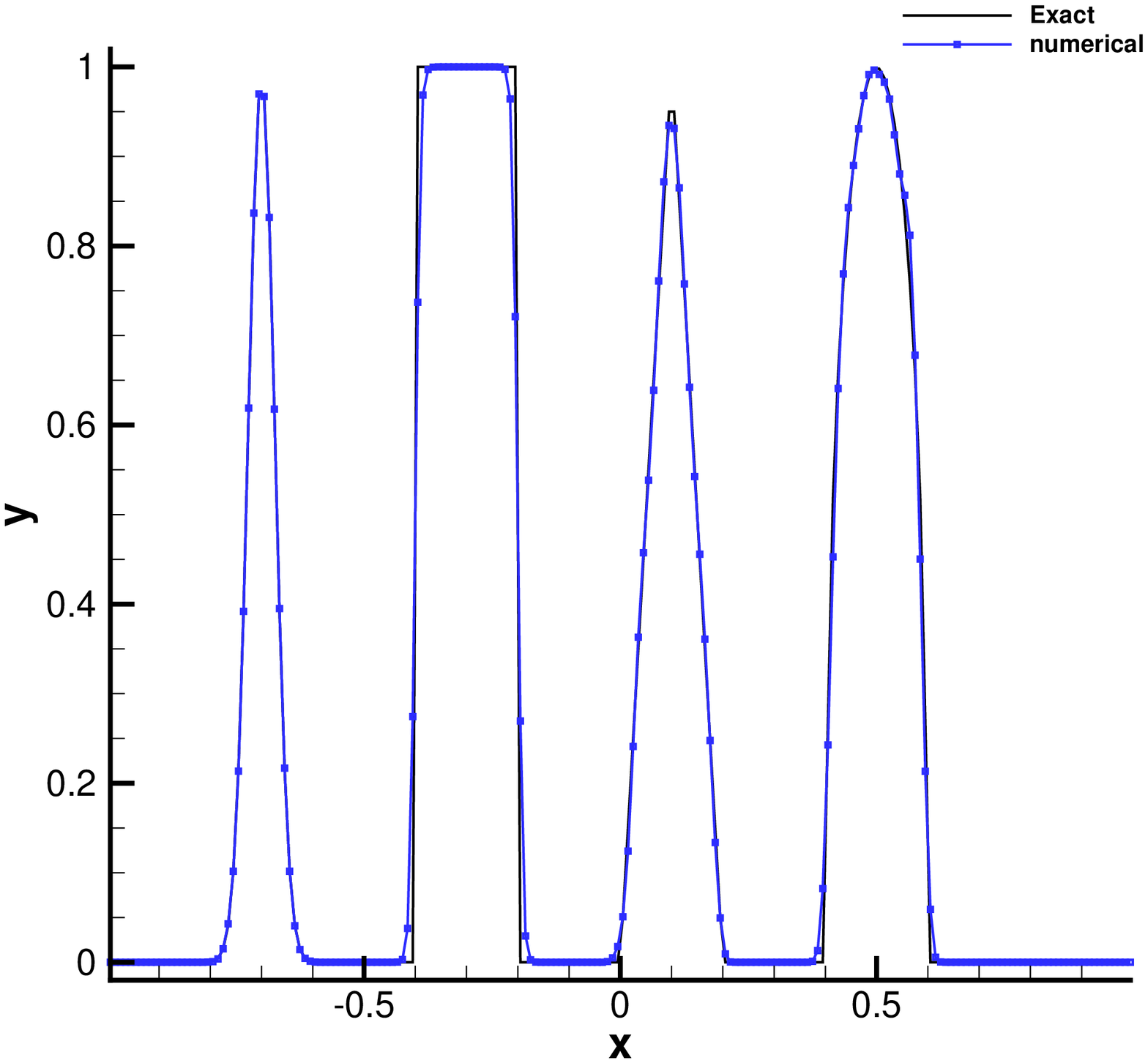}}
    \end{center}
	\protect\caption{Numerical results for advection of complex waves. The numerical solutions at $t=2.0$ with 200 mesh cells are presented.
		\label{fig:shujiang}}	
\end{figure}

Shock capturing schemes should be able to solve profiles of different smoothness with high resolutions as well as without numerical oscillations. In this subsection, we simulated the propagation of a complex wave \cite{7}. The initial profile contains both discontinuities and smooth regions with different smoothness, which is given by 
\begin{equation}
q\left(x\right)=\left\{
\begin{array}{lllll}
\frac{1}{6} \left[G\left(x,\beta,z- \delta\right)+G\left(x,\beta,z+\delta\right)+4G\left(x,\beta,z\right)\right]\ &\mathrm{if}\ \left|x+0.7\right| \leq 0.1,\\ %-0.8 \leq x \leq -0.6 \\
1\ &\mathrm{if}\ \left|x+0.3\right| \leq 0.1,\\ 
1-\lvert 10\left(x-0.1\right)\lvert\ &\mathrm{if}\ \left|x-0.1\right| \leq 0.1,\\ 
\frac{1}{6} \left[F\left(x,\alpha,a- \delta\right)+F\left(x,\alpha,a+\delta\right)+4F\left(x,\alpha,a\right)\right)]\ &\mathrm{if}\ \left|x-0.5\right| \leq 0.1,\\ 
0\ &\mathrm{otherwise},
\end{array}
\right.,
\end{equation}
where functions $F$ and $G$ are defined as
\begin{equation}
G\left(x,\beta,z\right)=\exp\left[-\beta\left(x-z\right)^{2}\right],\ F\left(x,\alpha,a\right)=\sqrt{\max\left[1-\alpha^{2}\left(x-a\right)^{2},0\right]},
\end{equation}
and the coefficients are
\begin{equation}
a=0.5,\ z=0.7,\ \delta=0.005,\ \alpha=10.0,\ \beta=\mathrm{log}_2\left(36\delta^{2}\right).
\end{equation}
The computation was carried out for one period at $t=2.0$ with a 200-cell mesh. The results calculated by the $\mathrm{P}_{n}\mathrm{T}_{m}-\mathrm{BVD-CD}$ schemes with different order were presented in Fig.~\ref{fig:shujiang}. It can be seen that all of schemes are essentially oscillation free and capture sharper discontinuities, which is important to solve shear layer, contact plan and material interface with high fidelity. With polynomial degree increased, the extreme points of the initial profile are better resolved . Compared with upwind-biased scheme \cite{41}, the central scheme shows advantageous in solving extremes. It is also noteworthy that $\mathrm{P}_{n}\mathrm{T}_{m}-\mathrm{BVD-CD}$ schemes produce almost same results for the discontinuity which is resolved by only four cells since the BVD-CD algorithm can properly choose THINC reconstruction function across discontinuities. Compared with other central-upwind shock-capturing scheme (see the Fig.~4 in \cite{LiChen}), the present solution is one of best with given cell elements.     

\subsection{Sod's problem}
\begin{figure}
	\begin{center}
	\subfigure[6th order scheme]
	{\centering\includegraphics[width=.32\textwidth,trim={0.5cm 0.5cm 0.5cm 0.5cm},clip]{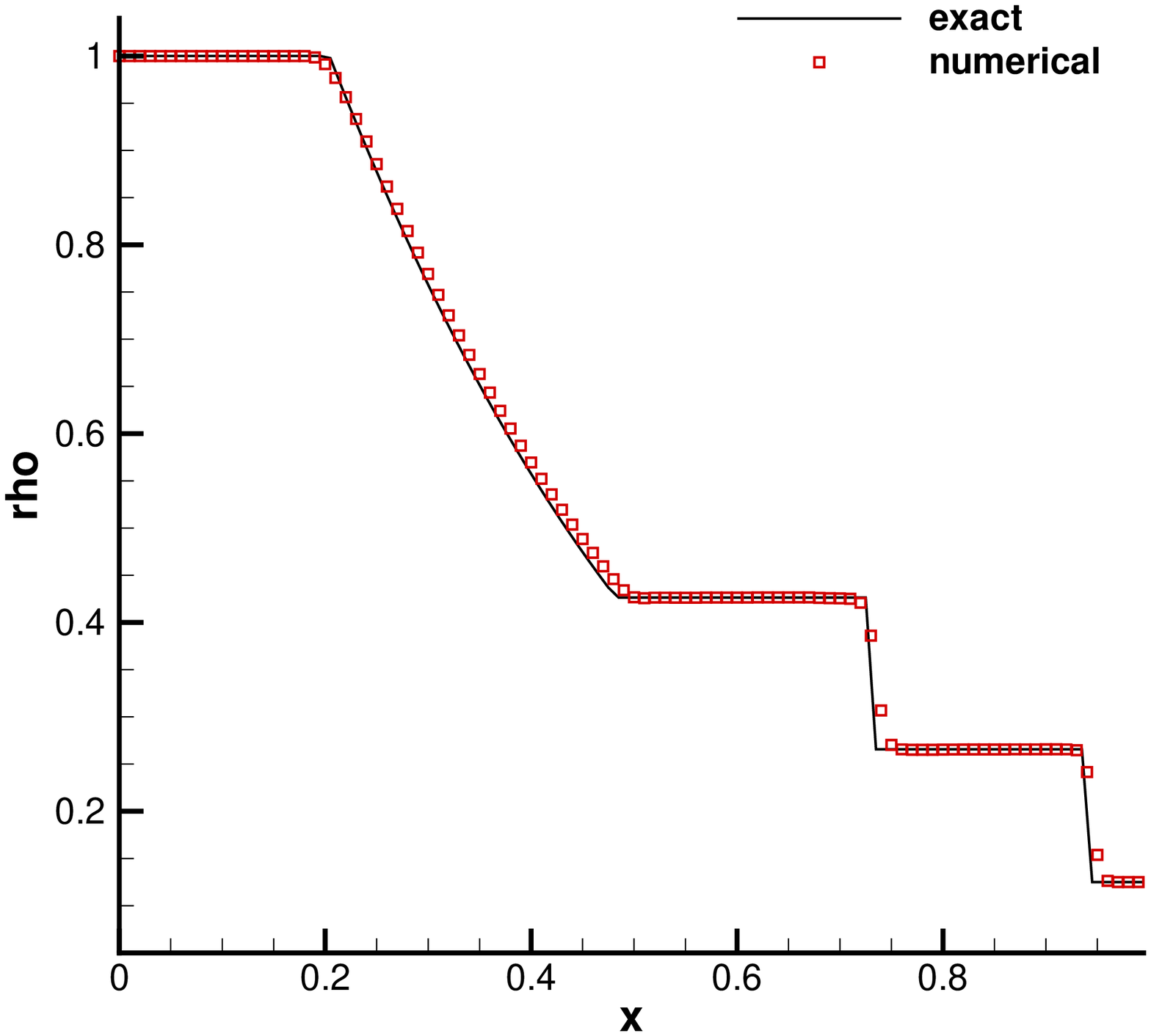}}
    \subfigure[8th order scheme]
	{\centering\includegraphics[width=.32\textwidth,trim={0.5cm 0.5cm 0.5cm 0.5cm},clip]{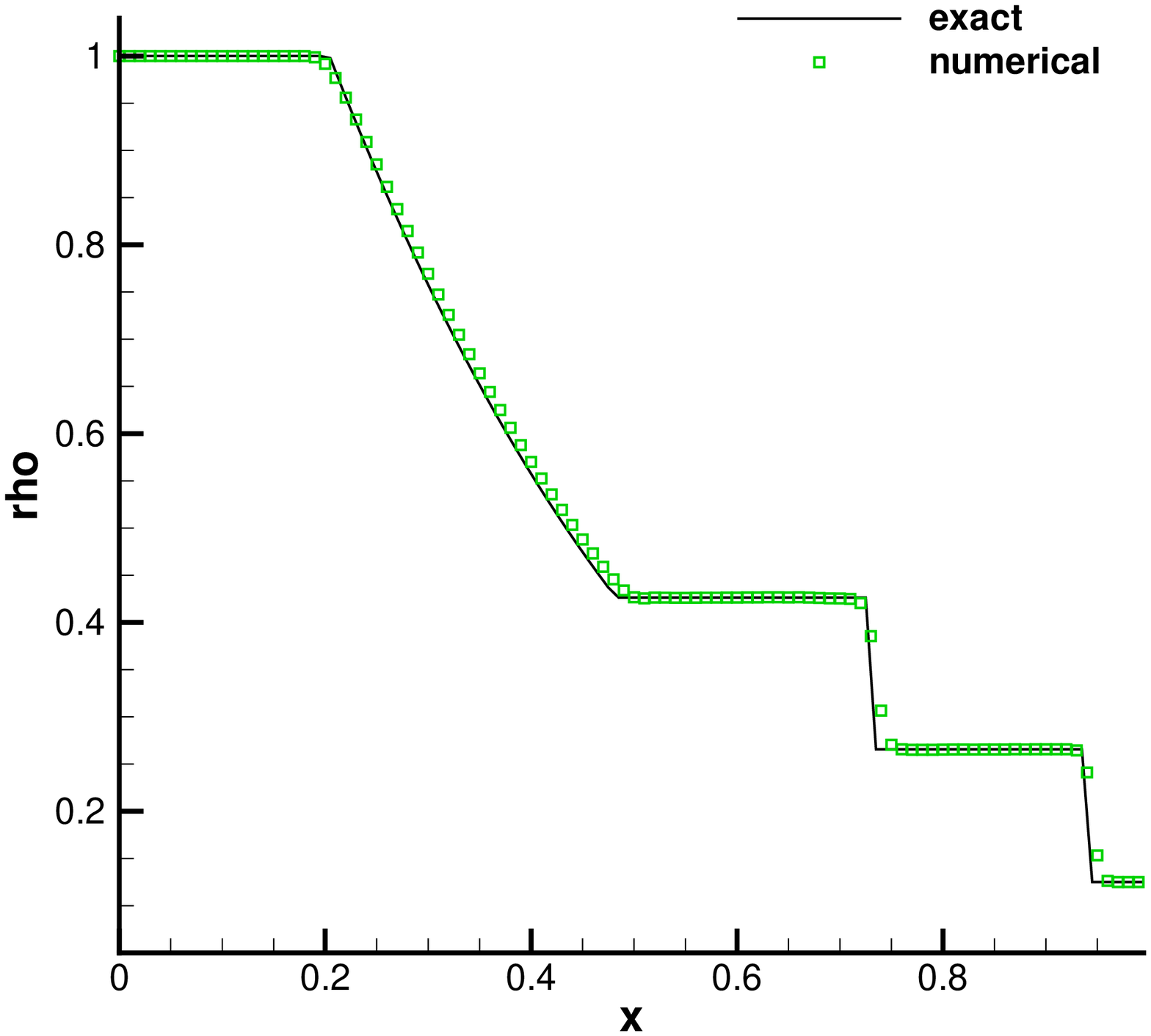}}
	\subfigure[10th order scheme]
	{\centering\includegraphics[width=.32\textwidth,trim={0.5cm 0.5cm 0.5cm 0.5cm},clip]{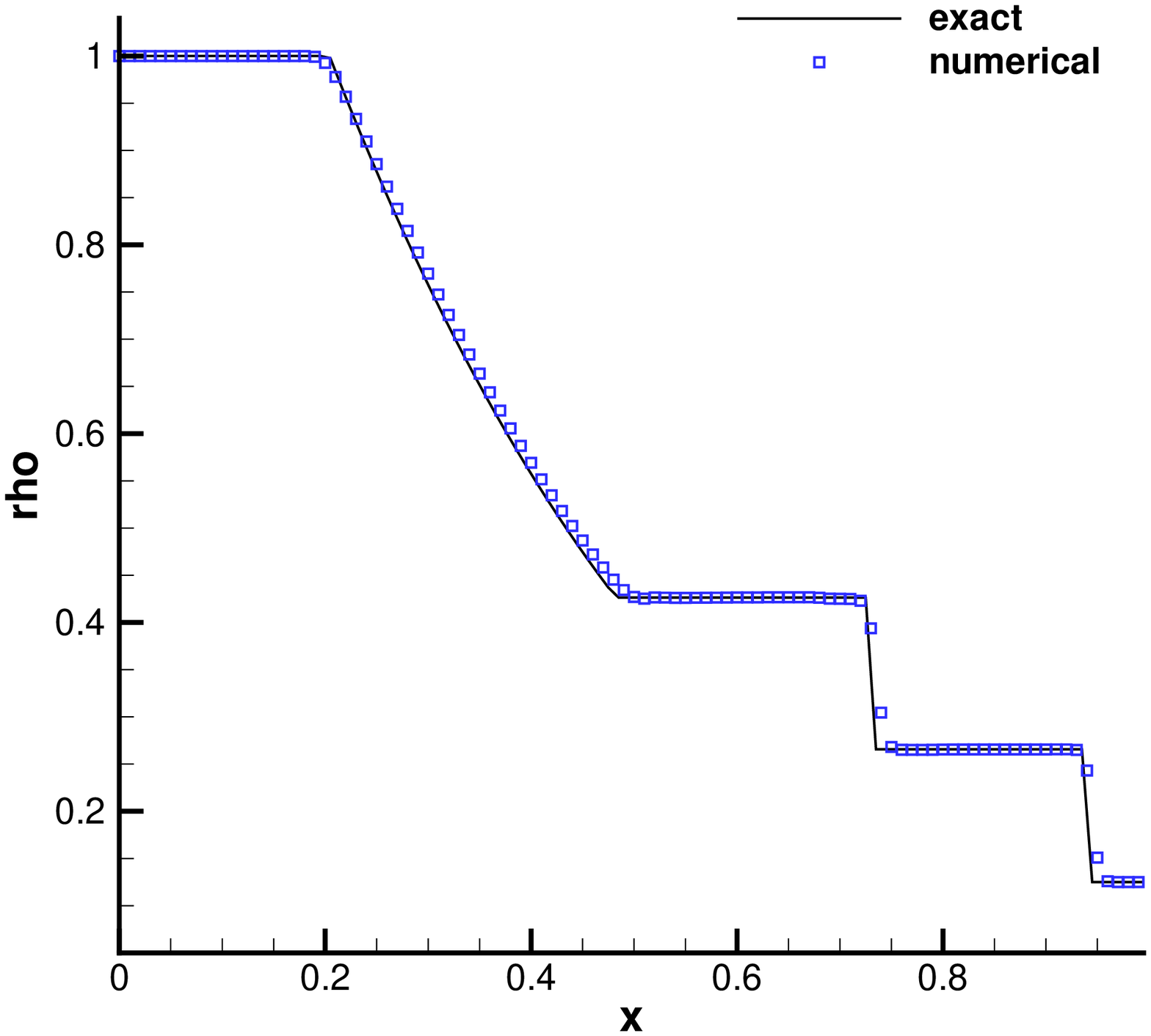}}
	\end{center}
	\protect\caption{Numerical results of Sod's problem for density field at $t = 0.25$ with $100$ cells. 
    \label{fig:sodrho}}	
\end{figure}

The Sod's problem is employed here to test the performance of present schemes in solving Euler equation. The initial distribution on computational domain $[0,1]$ was specified as \cite{sod}
\begin{equation}
\left(\rho_{0},\ u_{0},\ p_{0}\right)=\left\{
\begin{array}{ll}
\left(1,\ 0,\ 1\right) & 0 \leq x \leq 0.5 \\
\left(0.125,\ 0,\ 0.1\right) & \mathrm{otherwise}
\end{array}
\right..
\end{equation}
The computation was carried out on a mesh of  100 uniform cells up to $t=0.25$. The numerical results calculated from the proposed scheme were shown in Fig.~\ref{fig:sodrho} for density fields. We observe that $\mathrm{P}_{n}\mathrm{T}_{m}-\mathrm{BVD-CD}$ schemes can solve the contact discontinuity and right moving shock waves without obvious numerical oscillations. Moreover, the contact is resolved within only two cells. As expected, $\mathrm{P}_{n}\mathrm{T}_{m}-\mathrm{BVD-CD}$ schemes of different orders produce similar results across the contact because the THINC function is selected by the BVD-CD algorithm. 

\subsection{Lax's problem}
\begin{figure}
	\begin{center}
	\subfigure[6th order scheme]
	{\centering\includegraphics[width=.32\textwidth,trim={0.5cm 0.5cm 0.5cm 0.5cm},clip]{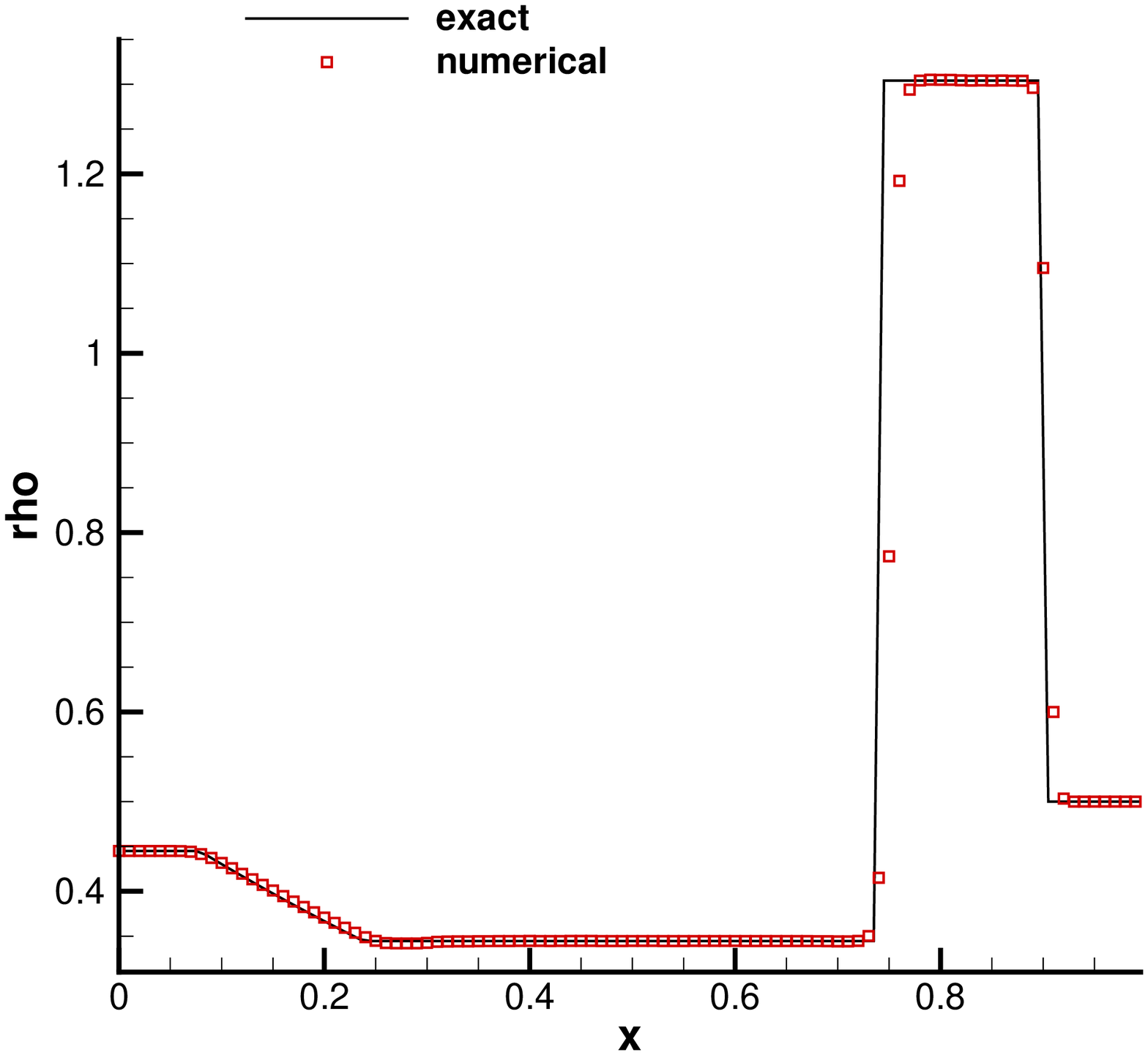}}
    \subfigure[8th order scheme]
	{\centering\includegraphics[width=.32\textwidth,trim={0.5cm 0.5cm 0.5cm 0.5cm},clip]{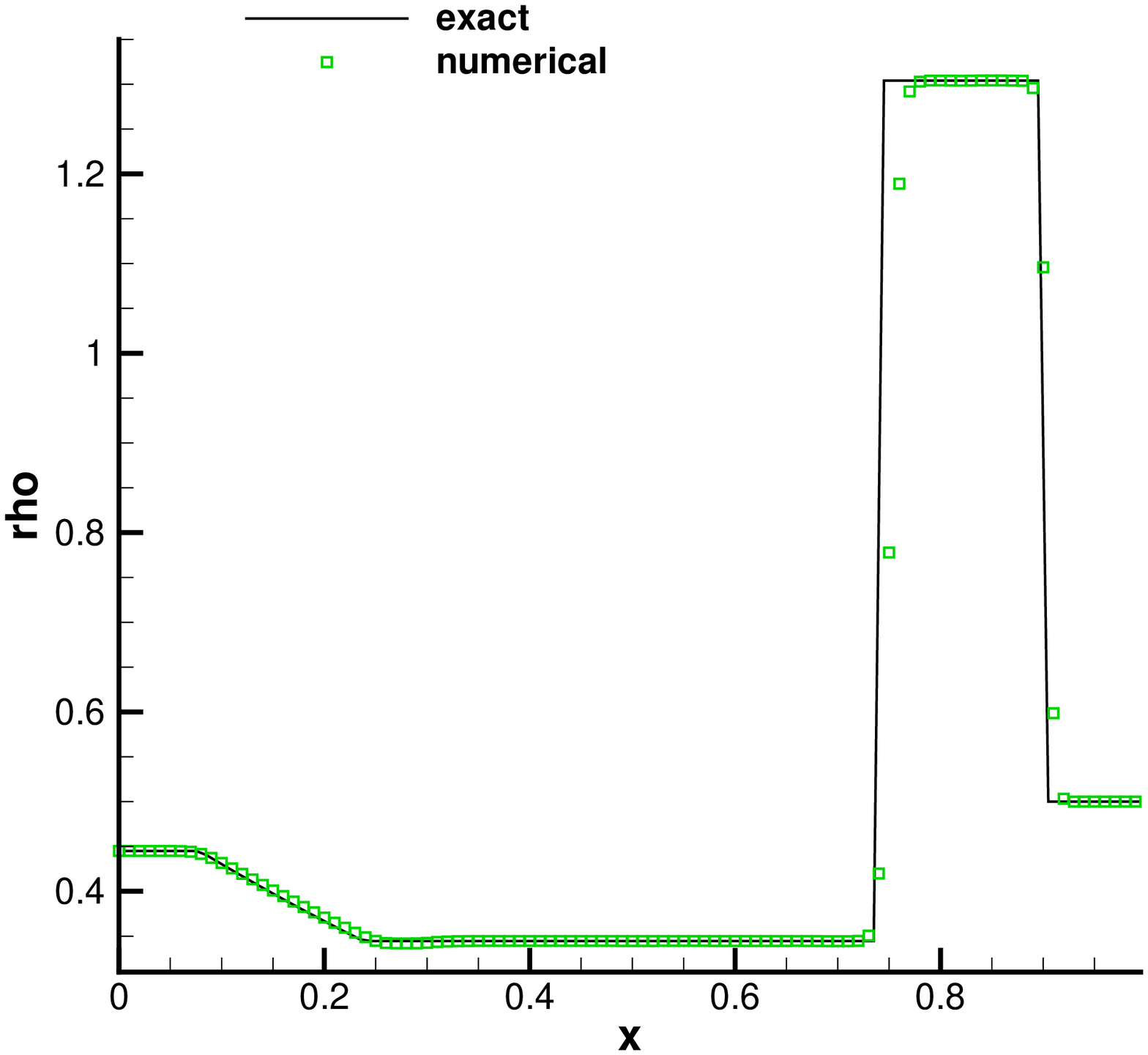}}
	\subfigure[10th order scheme]
	{\centering\includegraphics[width=.32\textwidth,trim={0.5cm 0.5cm 0.5cm 0.5cm},clip]{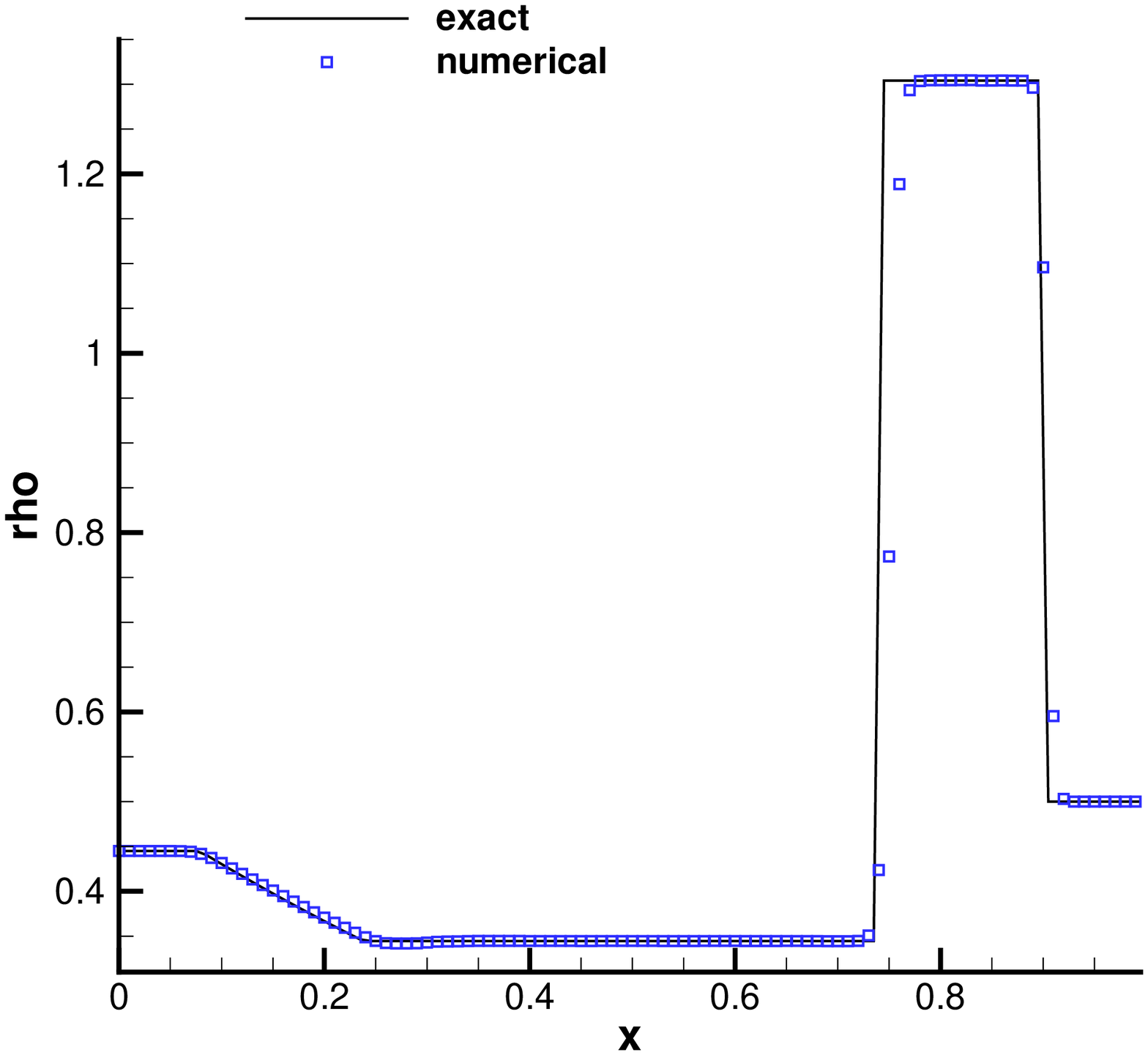}}
  \end{center}
  \protect\caption{Numerical results of Lax's problem for density field at time $t = 0.16$ with $100$ cells. 
    \label{fig:lax}}	
\end{figure}

We solve the Lax problem \cite{6} which contains relatively strong shock in this subsection. The initial condition is given by
\begin{equation}
\left(\rho_{0},\ u_{0},\ p_{0}\right)=\left\{
\begin{array}{ll}
\left(0.445,\ 0.698,\ 3.528\right) & 0 \leq x \leq0.5\\
\left(0.5,\ 0.0,\ 0.571\right) & \mathrm{otherwise}
\end{array}
\right..
\end{equation}
With the same number of cells as in the previous test case, we got the numerical results at $t=0.16$. The density field is plotted presented in Fig.~\ref{fig:lax}. The proposed $\mathrm{P}_{n}\mathrm{T}_{m}-\mathrm{BVD-CD}$ schemes solve sharp contact and shock waves without numerical oscillation. Compared with one of recently developed central-upwind schemes (Fig.~9 in \cite{LiChen}), the current scheme produces less oscillatory and less diffusive results. 

\subsection{Strong Lax's problem}
\begin{figure}
	\begin{center}
	\subfigure[6th order scheme]
	{\centering\includegraphics[width=.32\textwidth,trim={0.5cm 0.5cm 0.5cm 0.5cm},clip]{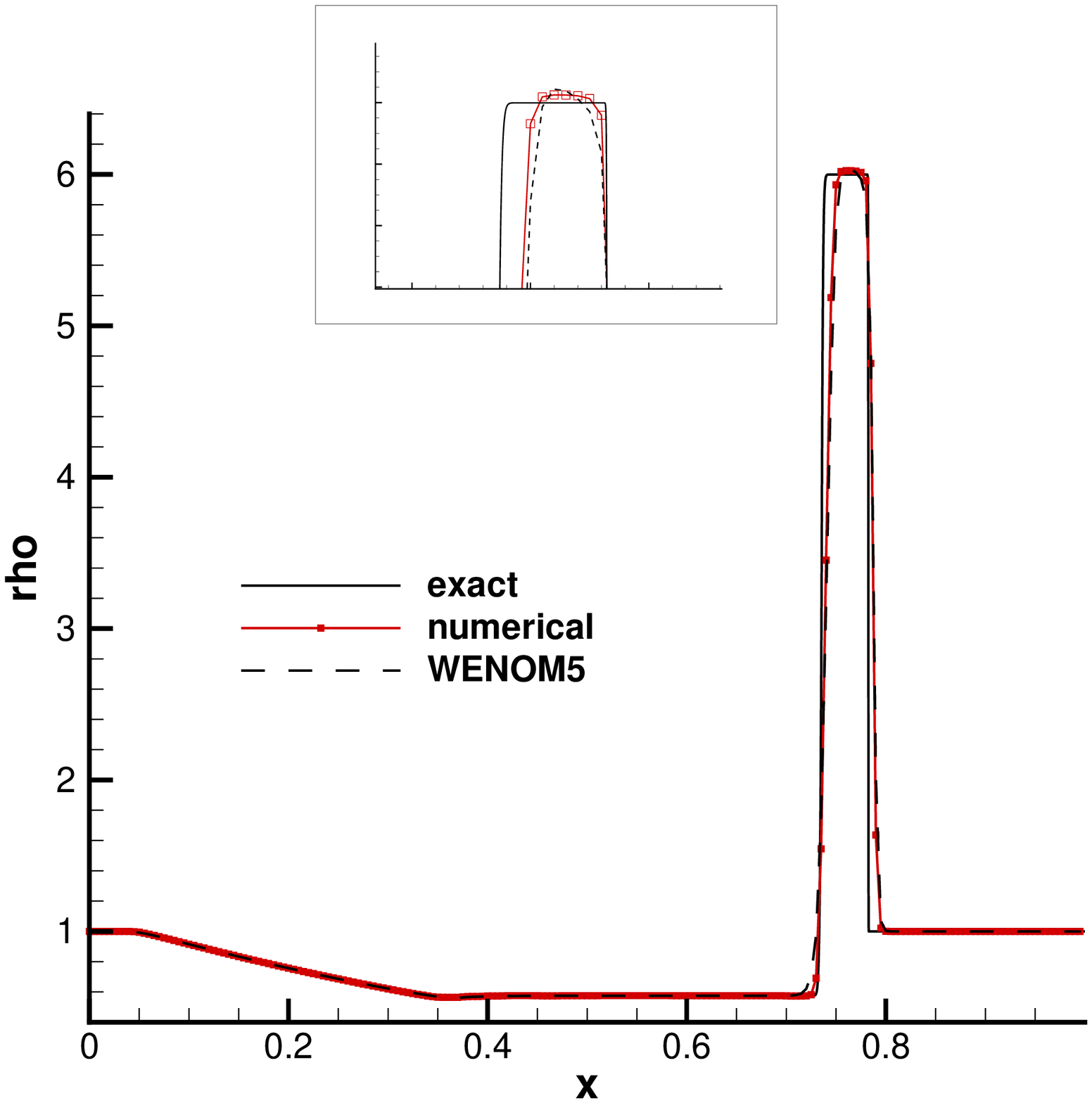}}
    \subfigure[8th order scheme]
	{\centering\includegraphics[width=.32\textwidth,trim={0.5cm 0.5cm 0.5cm 0.5cm},clip]{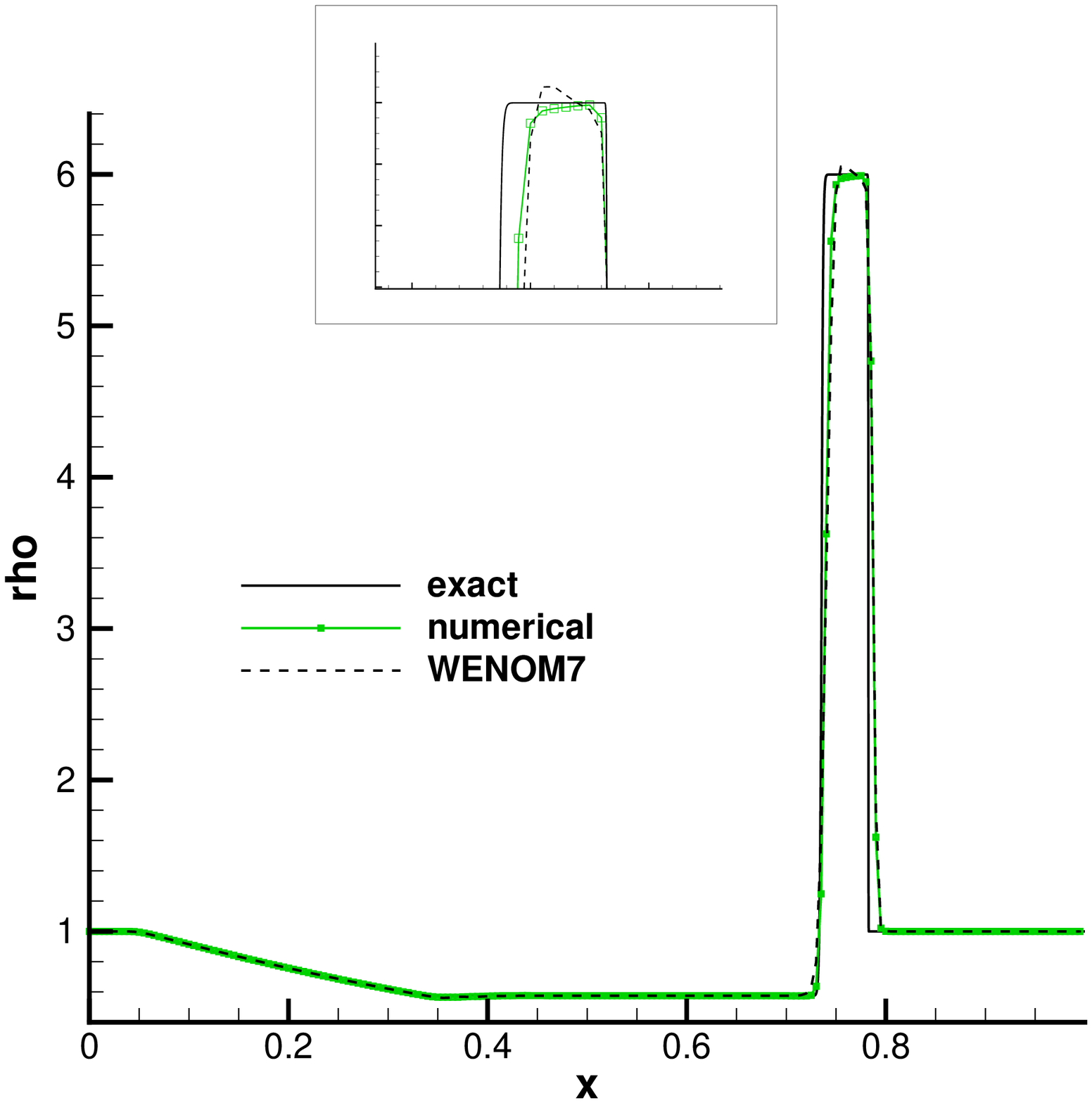}}
	\subfigure[10th order scheme]
	{\centering\includegraphics[width=.32\textwidth,trim={0.5cm 0.5cm 0.5cm 0.5cm},clip]{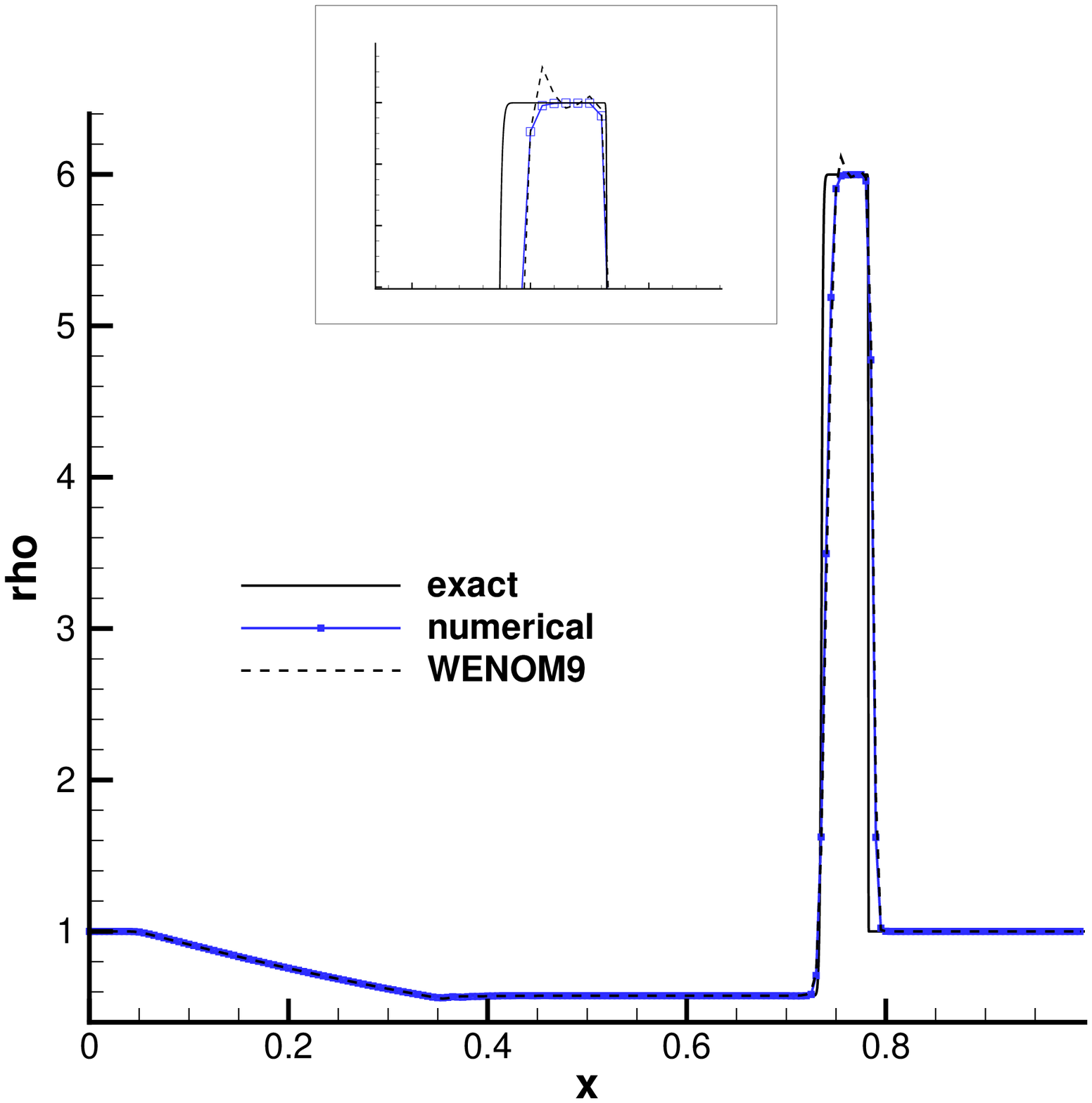}}
	\protect\caption{Numerical results of strong Lax's problem with Mach 198 for density field at time $t = 0.012$ with $200$ cells. Comparisons are made between WENOM and $\mathrm{P}_{n}\mathrm{T}_{m}-\mathrm{BVD-CD}$ at the same order. 
		\label{fig:stronglax}}	
	\end{center}
\end{figure}

Dealing with high Mach number flow involving strong shock is very challenging for low-dissipative schemes. Here another Lax's problem with strong discontinuities is used to test the robustness of different schemes on shock capturing. The initial condition is prescribed as
\begin{equation}
\left(\rho_{0},\ u_{0},\ p_{0}\right)=\left\{
\begin{array}{ll}
\left(1.0,\ 0.0,\ 1000.0\right), & 0 \leq x \leq0.5\\
\left(1.0,\ 0.0,\ 0.01\right), & \mathrm{otherwise}
\end{array}
\right..
\end{equation}
With initial high pressure ratio, a right-moving Mach 198 shock and a contact is generated. The computation lasts until time $t=0.012$ with 200 cell elements. The numerical solutions from different schemes for density field are plotted in Fig.~\ref{fig:stronglax}. The zoomed regions between contact and strong shock are also presented in the figure. We compare $n+2$ order $\mathrm{P}_{n}\mathrm{T}_{m}-\mathrm{BVD-CD}$ with $n+1$ order WENOM schemes. With increased order, the WENO scheme produces obvious overshooting and numerical oscillations. However, $\mathrm{P}_{n}\mathrm{T}_{m}-\mathrm{BVD-CD}$ schemes give more faithful solution with reduced dissipation and suppressed overshooting. This test shows the proposed scheme is able to capture strong shock robustly even though it recovers to central scheme in smooth region.

\subsection{Shock density wave interaction problem}
\begin{figure}
	\begin{center}
	\subfigure[6th order scheme]
	{\centering\includegraphics[width=.32\textwidth,trim={0.5cm 0.5cm 0.5cm 0.5cm},clip]{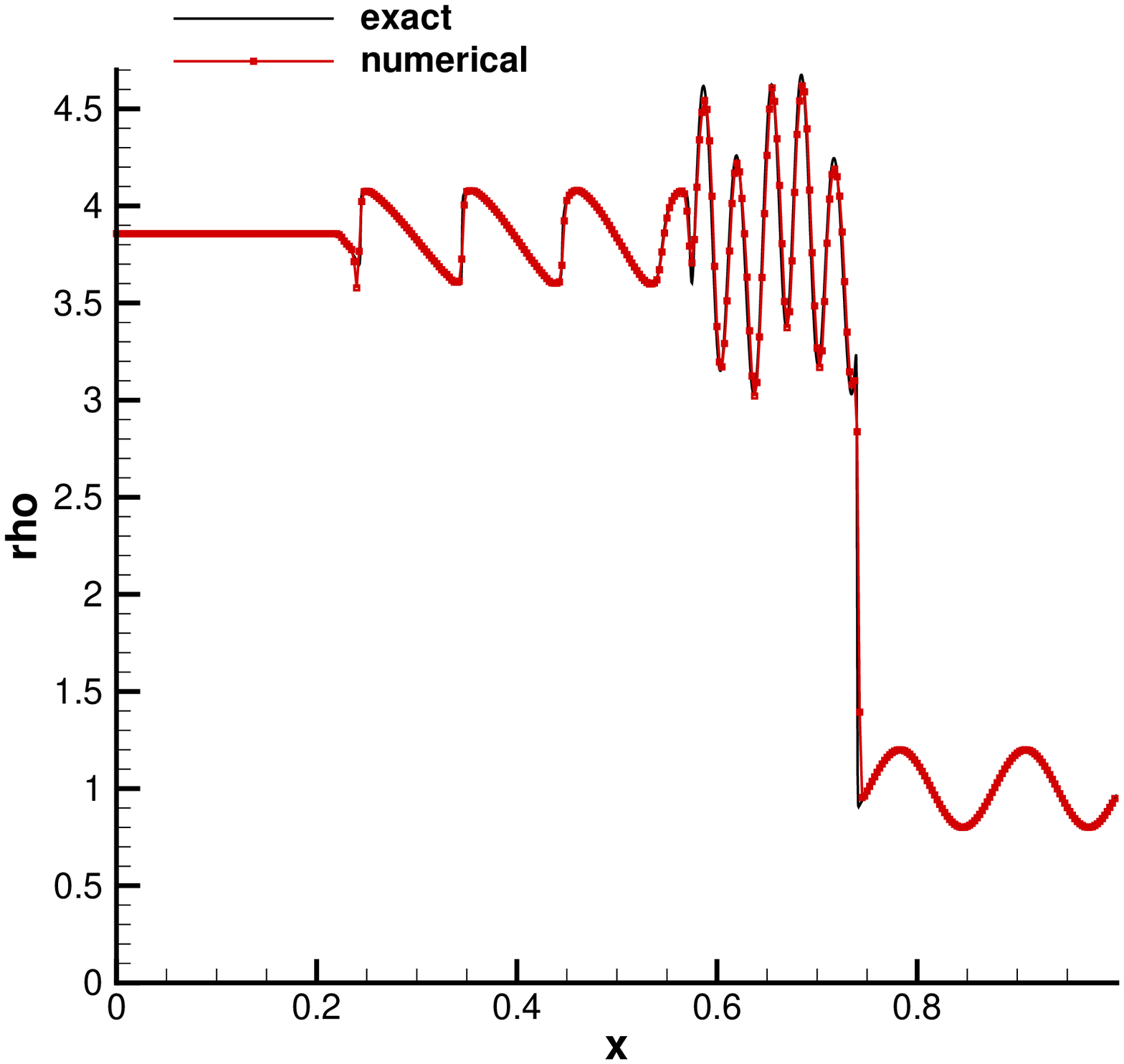}}
    \subfigure[8th order scheme]
	{\centering\includegraphics[width=.32\textwidth,trim={0.5cm 0.5cm 0.5cm 0.5cm},clip]{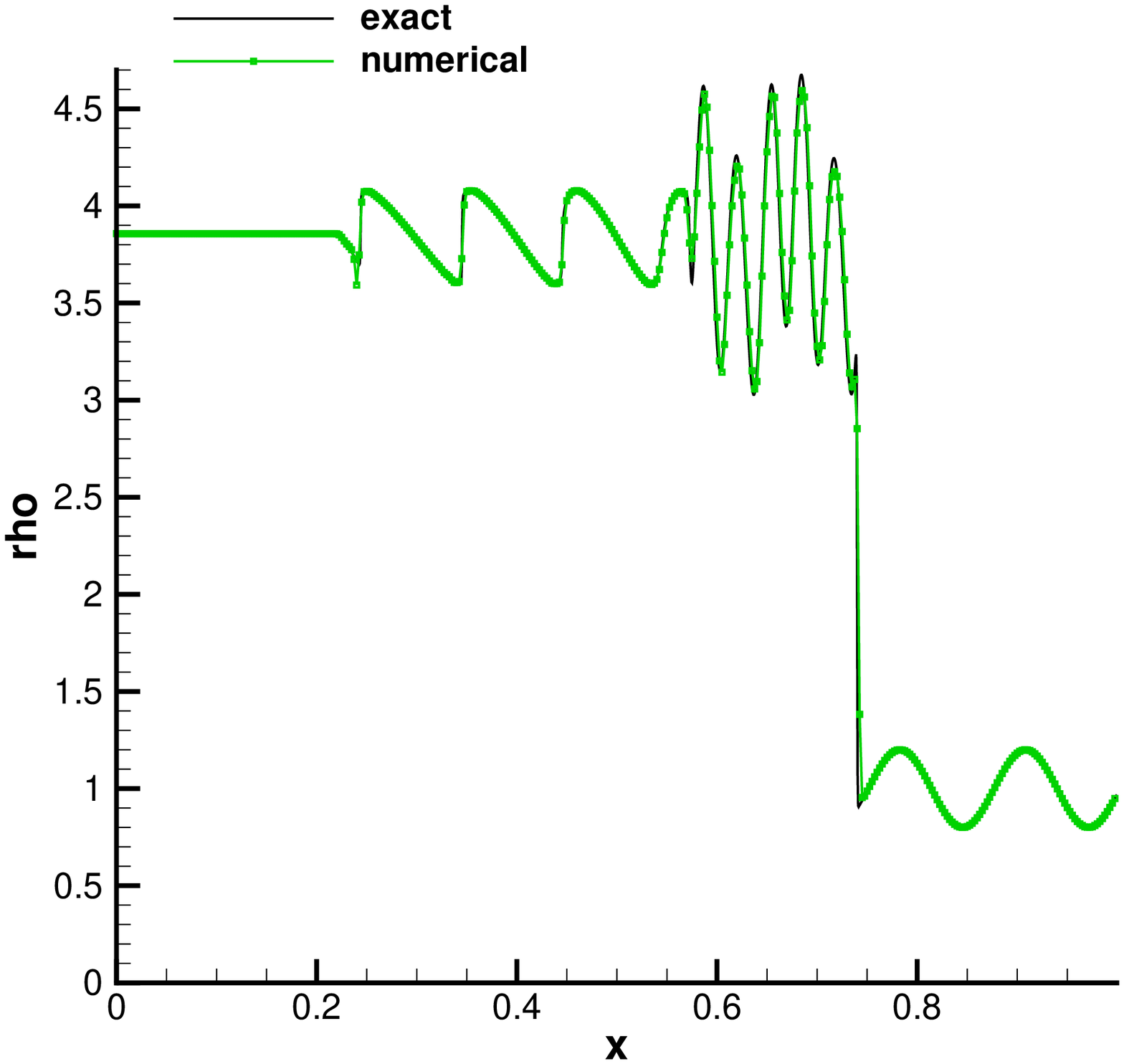}}
	\subfigure[10th order scheme]
	{\centering\includegraphics[width=.32\textwidth,trim={0.5cm 0.5cm 0.5cm 0.5cm},clip]{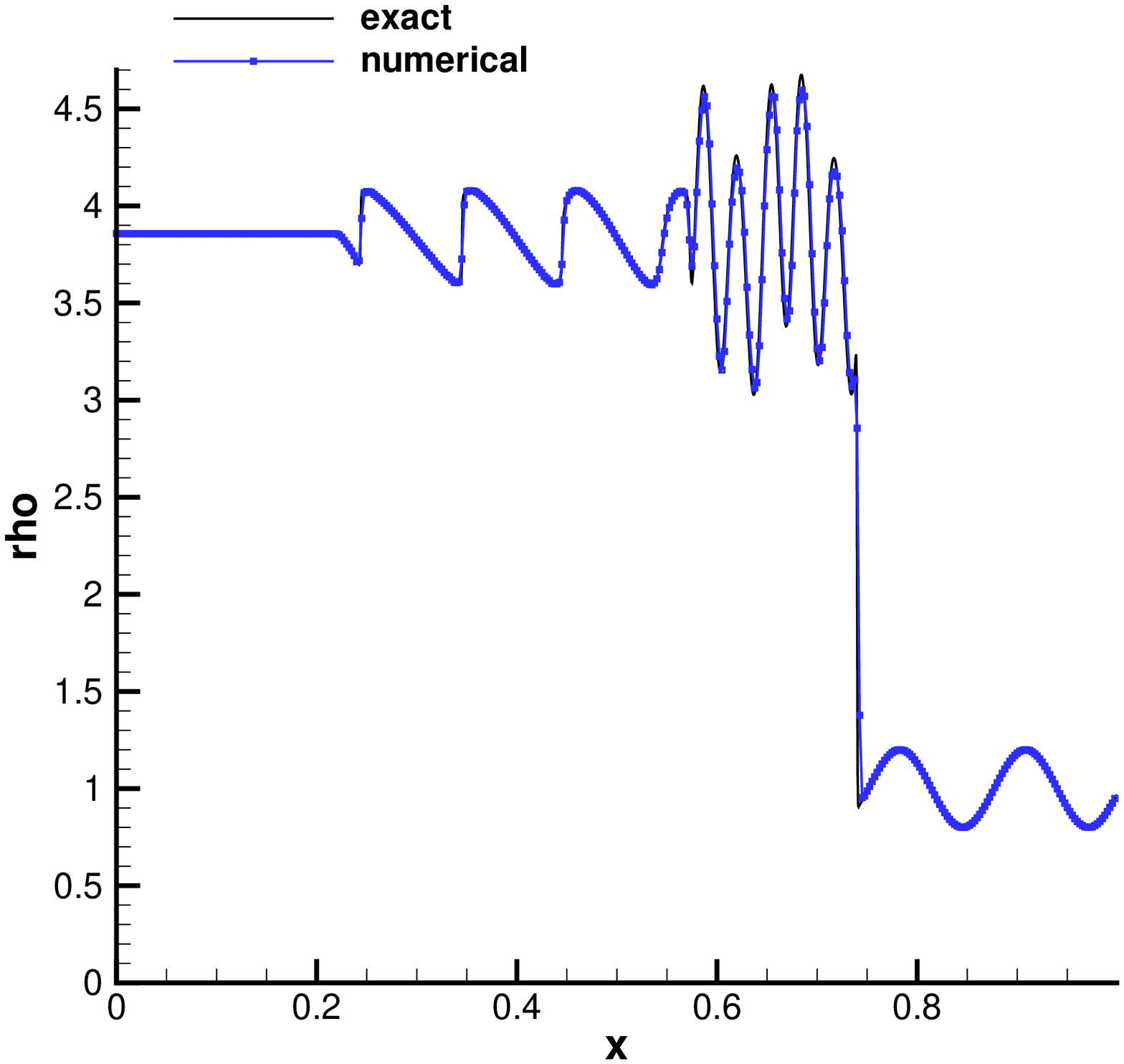}}
	\protect\caption{Numerical results of shock-density wave interaction problem at $t=0.18$ computed on 400 mesh elements. 
		\label{fig:shockT}}	
   \end{center}
\end{figure}

A Mach 4 shock wave interacting with a density disturbance is simulating with proposed schemes. The initial condition is set as
\begin{equation}
(\rho_{0},\ u_{0},\ p_{0})=\left\{
\begin{array}{ll}
\left(3.857148,\ 2.629369,\ 10.333333\right), \ &\mathrm{if}\ 0 \leq x \leq 0.1,\\
\left(1+0.2\sin\left(50x-25\right),\ 0,\ 1\right), \ &\mathrm{otherwise}.
\end{array}
\right.
\end{equation}
The numerical solutions at $t=0.18$ computed on 400 mesh elements are shown in Fig.~\ref{fig:shockT}, where the reference solution plotted by the solid line is computed by the classical 5th-order WENO scheme with 2000 mesh cells. The results show the proposed schemes are able to solve problem containing shocks and complex smooth flow features.

\begin{figure}
	\begin{center}
	\subfigure[6th order scheme]
	{\centering\includegraphics[width=.32\textwidth,trim={0.5cm 0.5cm 0.5cm 0.5cm},clip]{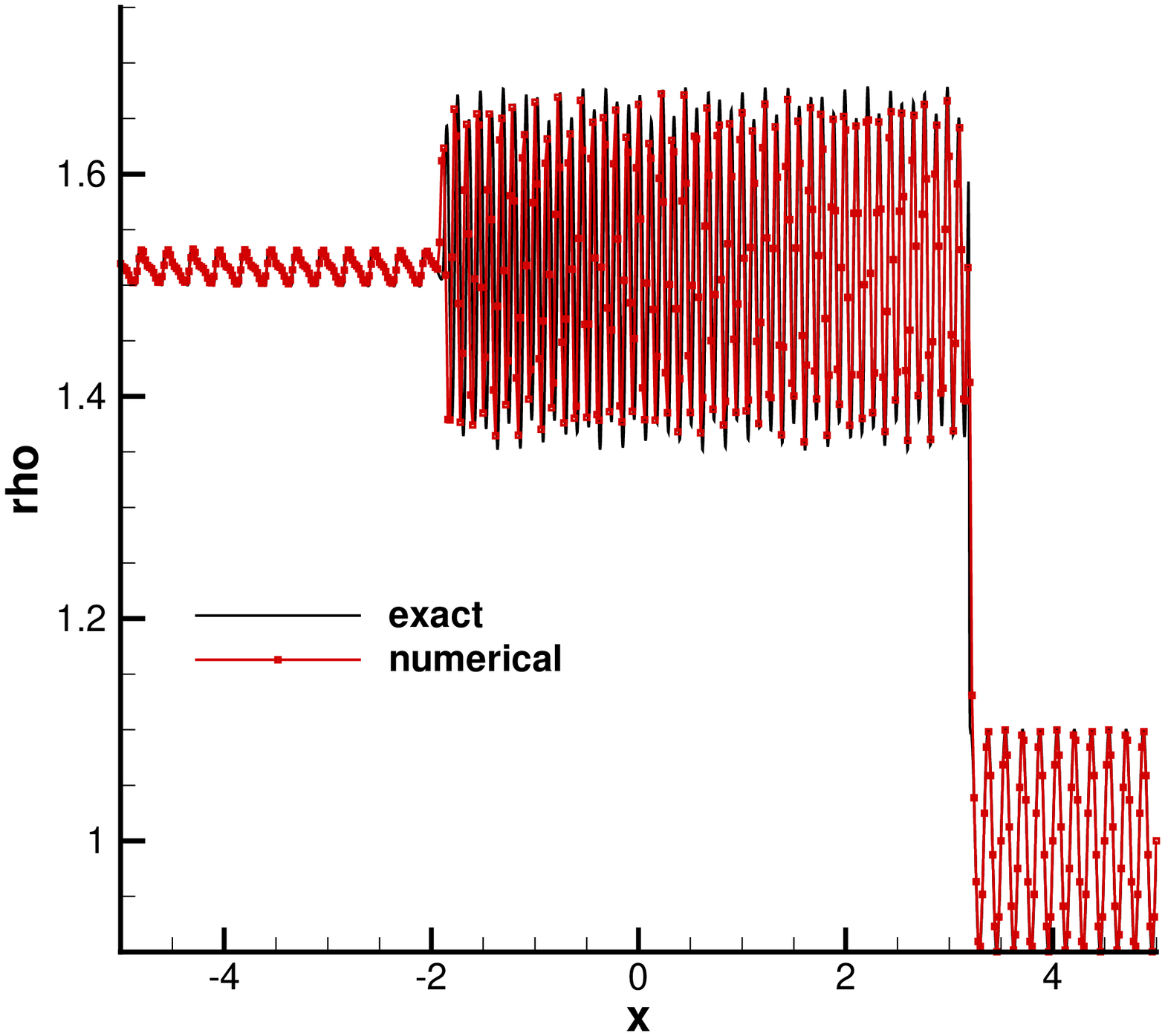}}
    \subfigure[8th order scheme]
	{\centering\includegraphics[width=.32\textwidth,trim={0.5cm 0.5cm 0.5cm 0.5cm},clip]{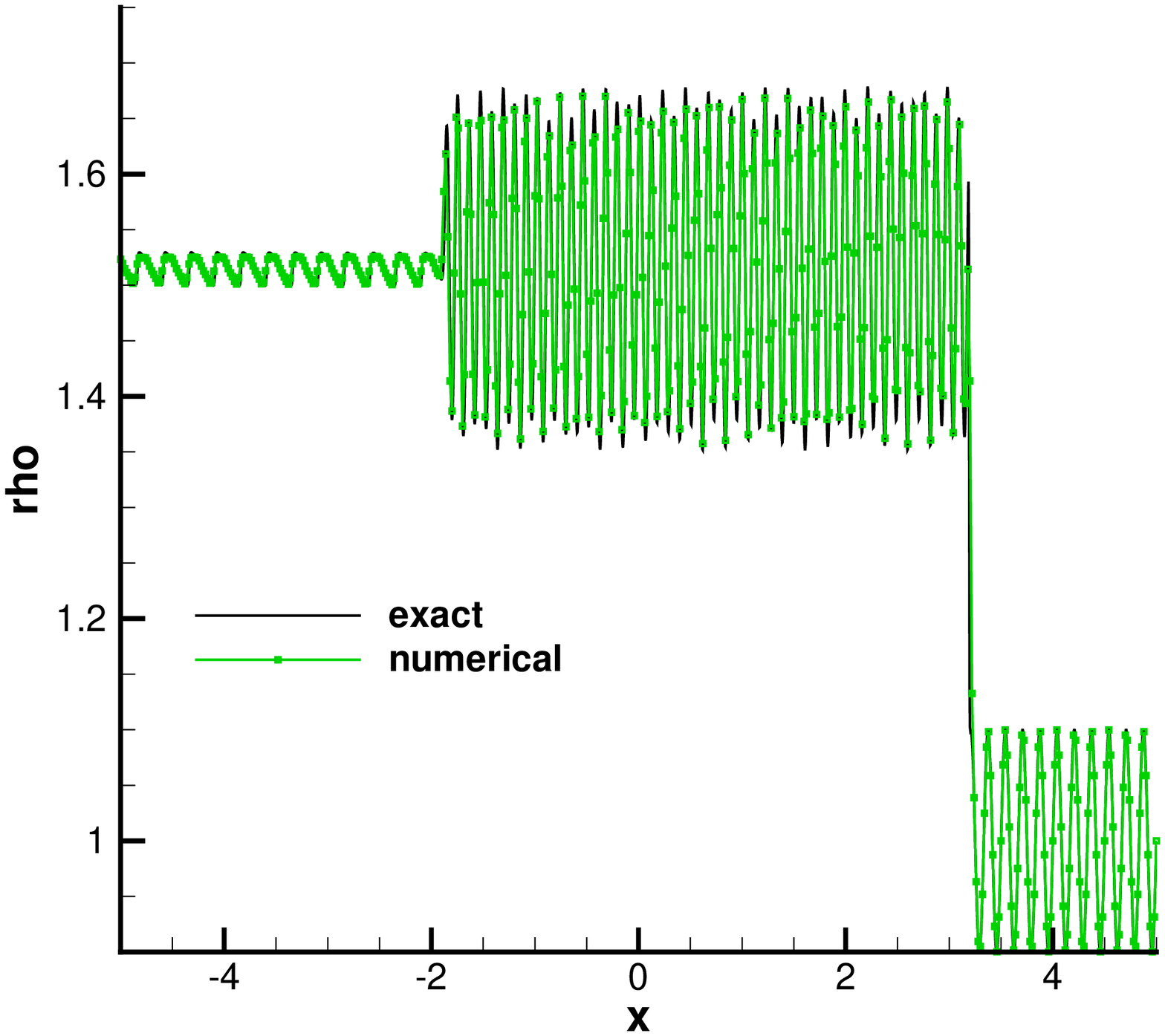}}
	\subfigure[10th order scheme]
	{\centering\includegraphics[width=.32\textwidth,trim={0.5cm 0.5cm 0.5cm 0.5cm},clip]{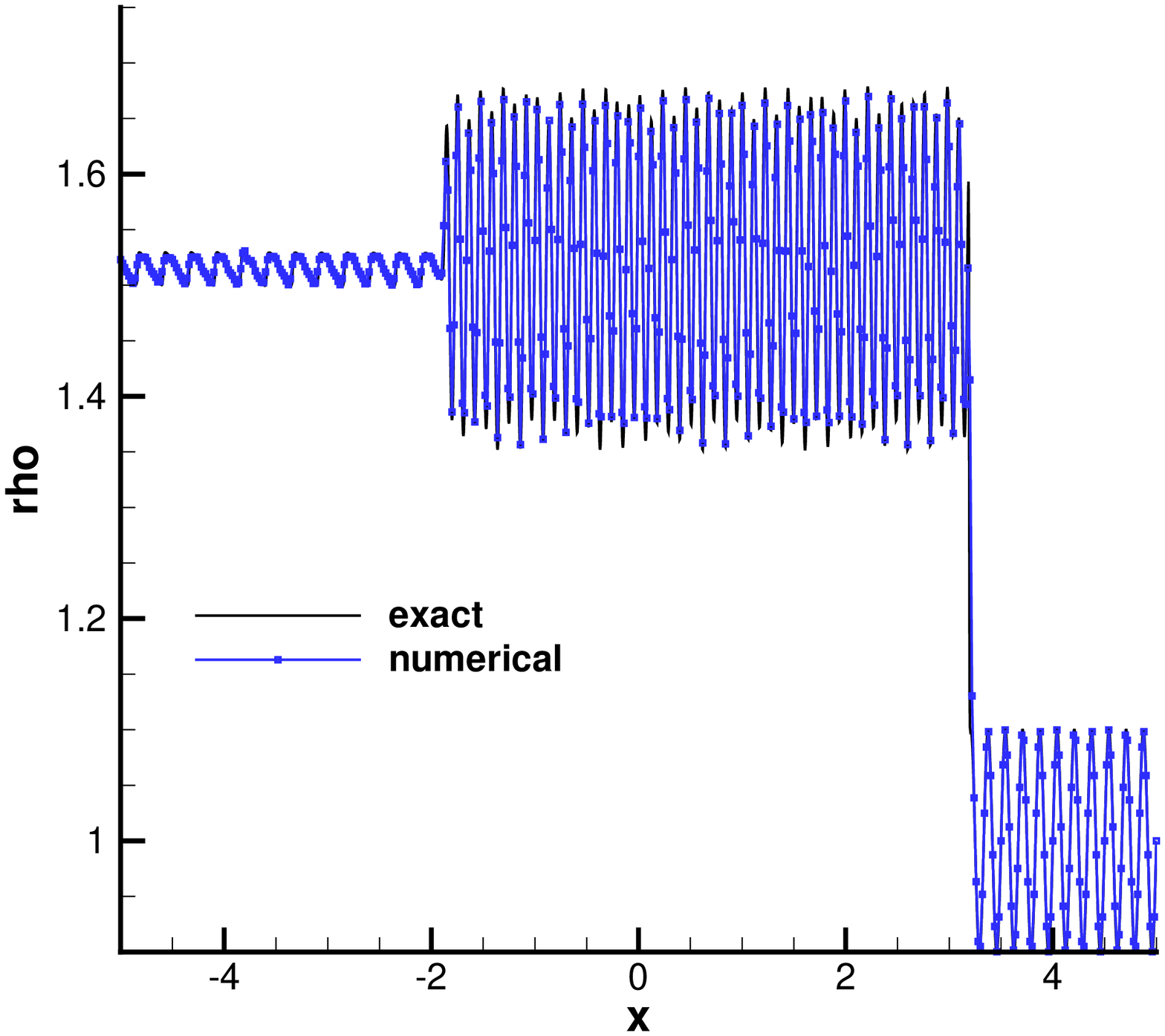}}
	\protect\caption{Numerical results of another shock-density wave interaction problem involving high wavenumber waves. Numerical results at time $t=5.0$ on 500 mesh elements are presented.
		\label{fig:TT}}	
   \end{center}
\end{figure}

\begin{figure}
	\begin{center}
	\subfigure[6th order scheme]
	{\centering\includegraphics[width=.32\textwidth,trim={0.5cm 0.5cm 0.5cm 0.5cm},clip]{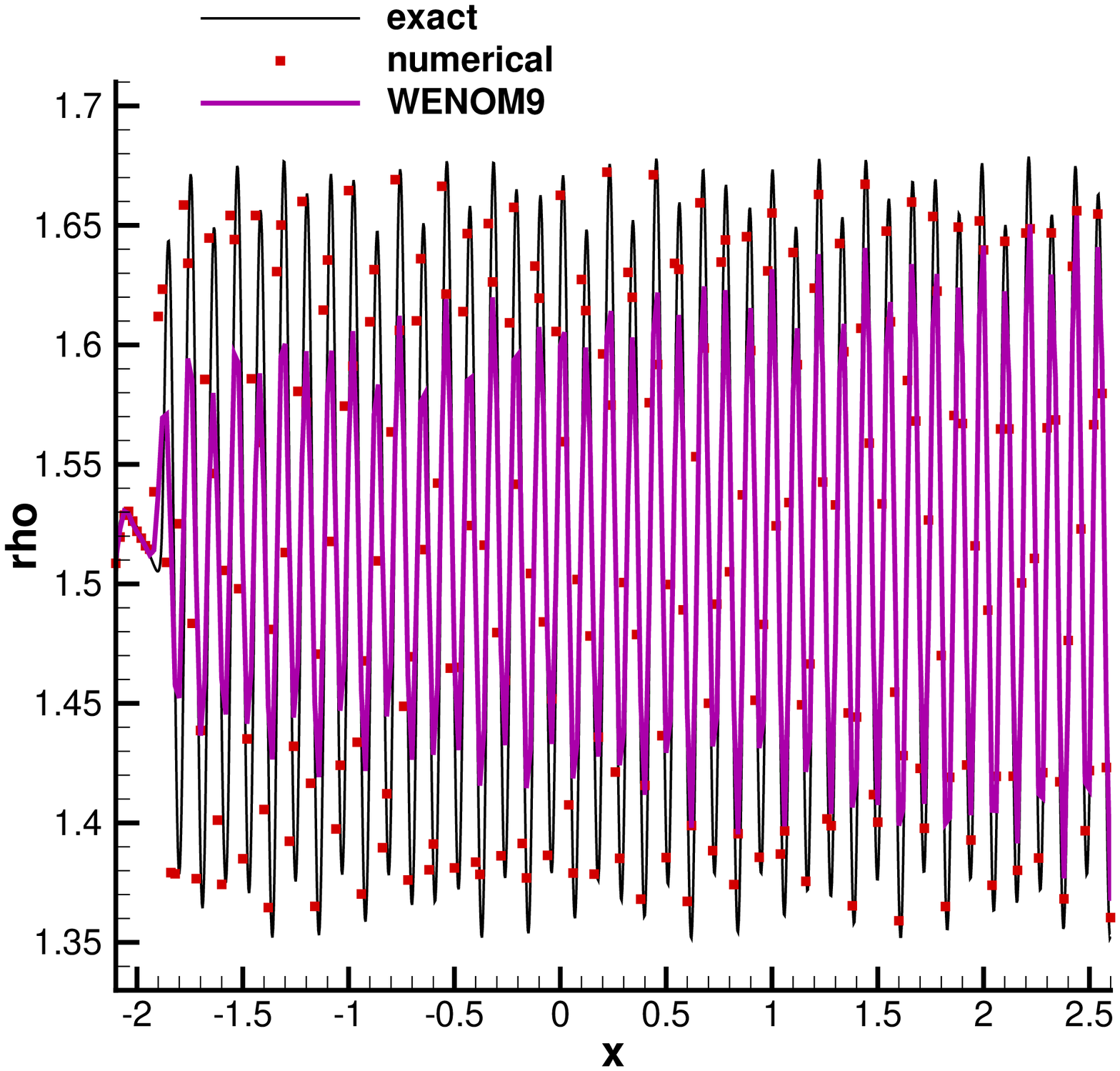}}
    \subfigure[8th order scheme]
	{\centering\includegraphics[width=.32\textwidth,trim={0.5cm 0.5cm 0.5cm 0.5cm},clip]{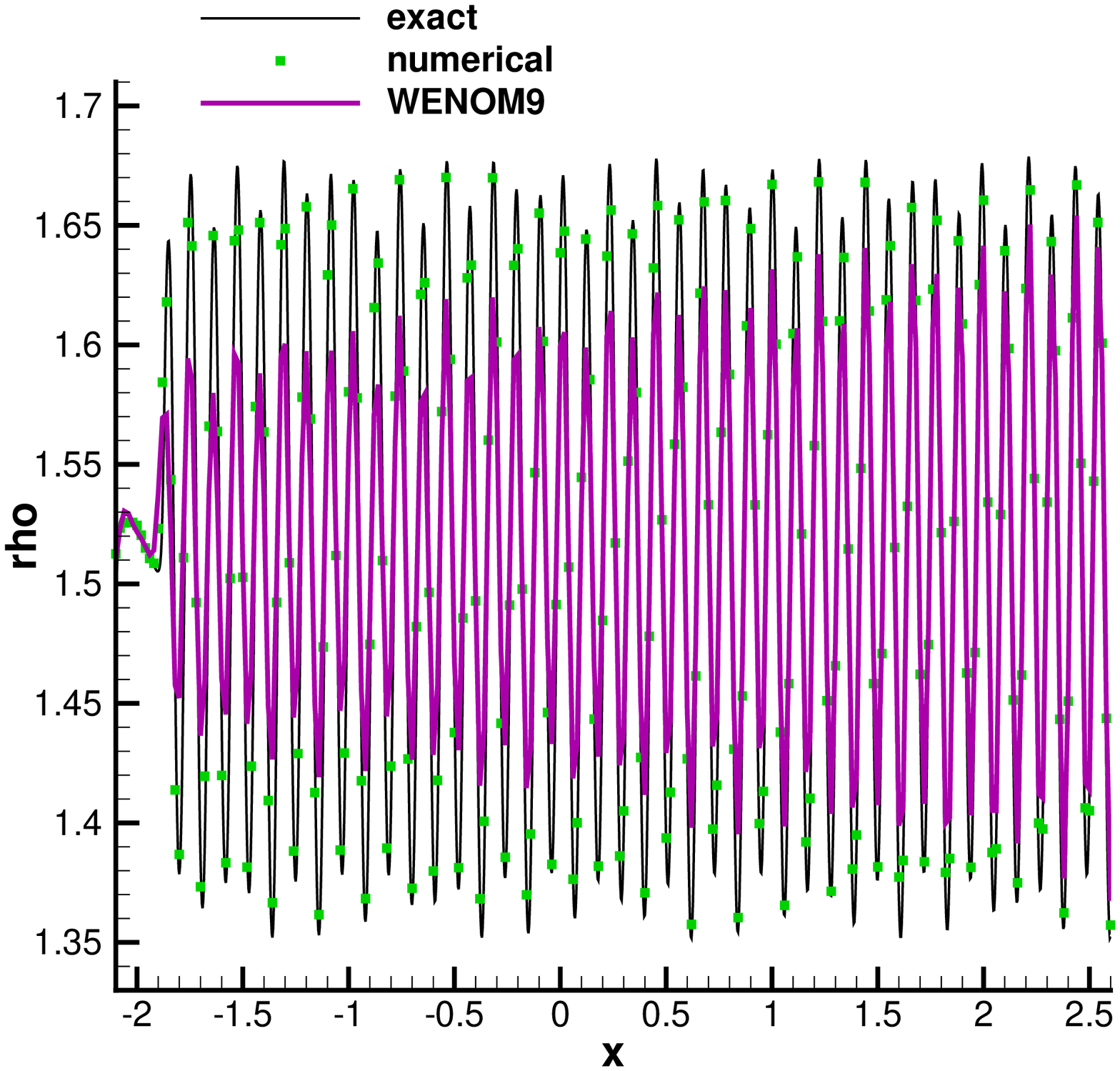}}
	\subfigure[10th order scheme]
	{\centering\includegraphics[width=.32\textwidth,trim={0.5cm 0.5cm 0.5cm 0.5cm},clip]{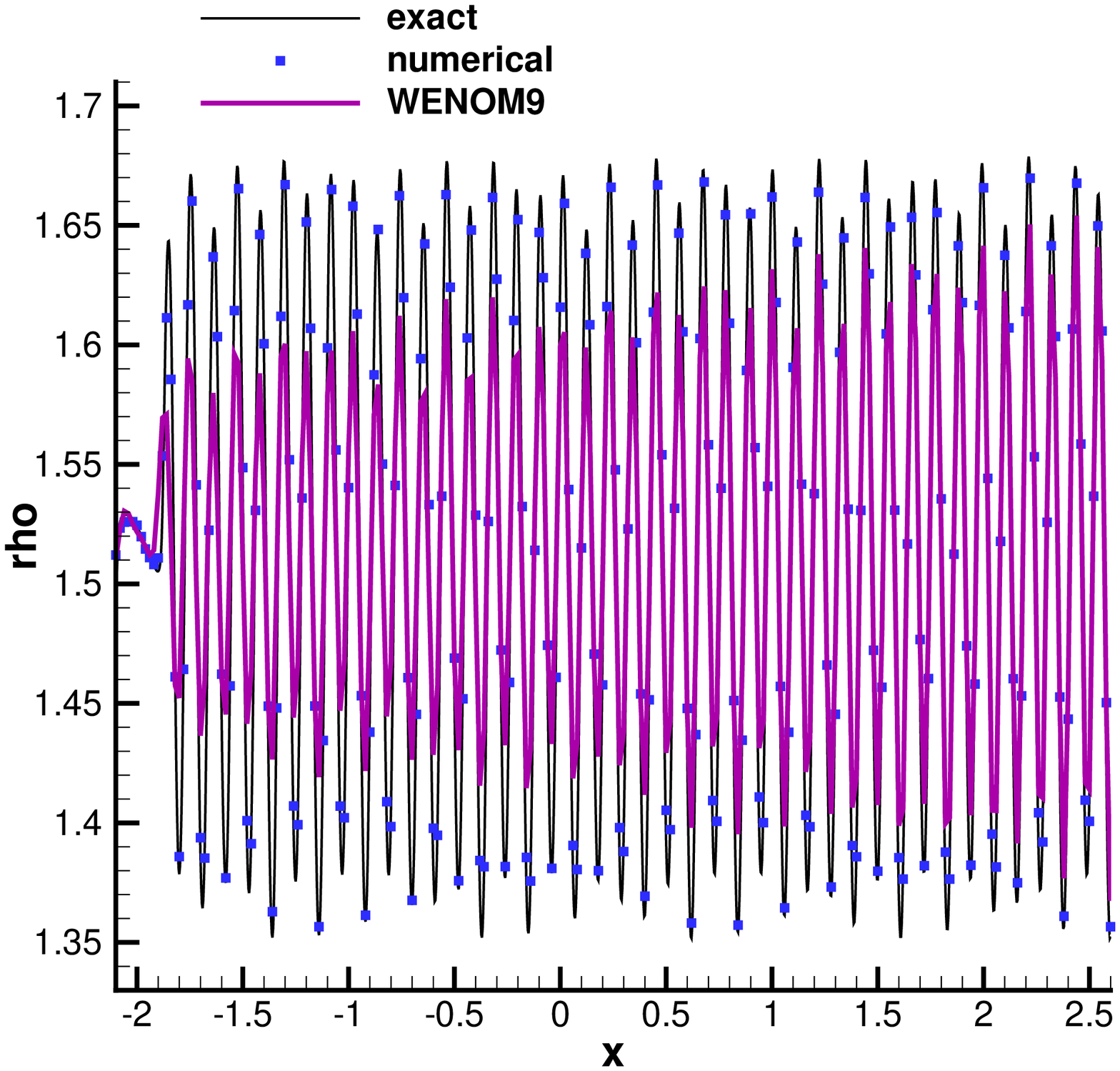}}
	\protect\caption{A zoomed region of the Fig.~\ref{fig:TT}. Comparisons were made between the $\mathrm{P}_{n}\mathrm{T}_{m}-\mathrm{BVD-CD}$ and the ninth order WENOM scheme.
		\label{fig:TTLL}}	
   \end{center}
\end{figure}

The advantage of introducing central flux is further demonstrated through another shock-density interaction problem involving waves of higher wavenumber. Similar to \cite{TitarevToro}, the initial condition is specified as 
\begin{equation}
(\rho_{0},\ u_{0},\ p_{0})=\left\{
\begin{array}{ll}
\left(1.515695,\ 0.523346,\ 1.805\right), \ &\mathrm{if}\  x \leq -4.5,\\
\left(1+0.1\sin(12 \pi x),\ 0,\ 1\right), \ &\mathrm{otherwise}.
\end{array}\right.
\end{equation}
The computation was carried out up to $t=5.0$ with 500 cells in a computational domain of $[-5.0,5.0]$. The numerical solutions are shown in Fig.~\ref{fig:TT}. It can been seen that the proposed schemes are able to resolve the complex structures involving shock and waves of high wavenumber with a relatively coarse mesh. The performance of the proposed schemes is further illustrated through comparing with the 9th order WENOM scheme in the Fig.~\ref{fig:TTLL} which presents a zoomed region for density perturbation. The results clearly show that the upwind-biased WENOM scheme is dissipative despite of its high order. On the contrary, the proposed scheme can capture the peak of waves through introducing non-dissipative central flux. In a summary, these 1D tests verify $\mathrm{P}_{n}\mathrm{T}_{m}-\mathrm{BVD-CD}$ schemes with $\lambda=0.5$ are able to resolve small-scale structures such as density perturbation of high wavenumbers. Also, the proposed schemes can solve discontinuous large-scale structures without producing obvious numerical oscillations.  

\subsection{2D viscous shock tube}
\begin{figure}
	\begin{center}
	\subfigure[$\mathrm{P_{4}T_{2}}$-BVD-CD]{\centering\includegraphics[scale=0.35,trim={0.5cm 0.5cm 0.5cm 0.5cm},clip]{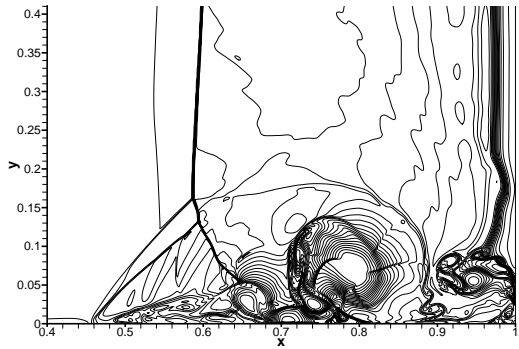}} 
	\subfigure[$\mathrm{P_{6}T_{3}}$-BVD-CD]{\centering\includegraphics[scale=0.35,trim={0.5cm 0.5cm 0.5cm 0.5cm},clip]{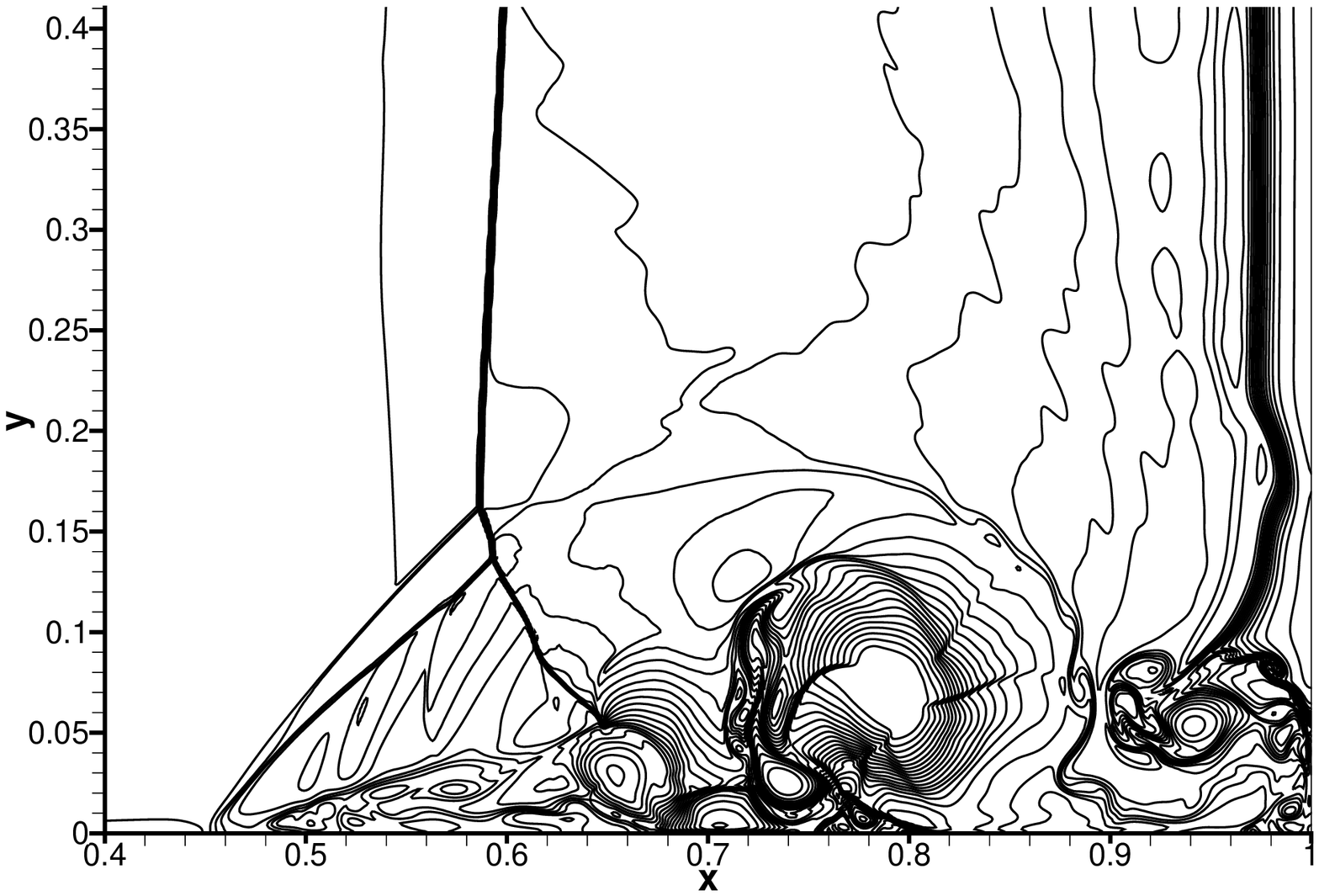}} 
	\subfigure[$\mathrm{P_{8}T_{3}}$-BVD-CD]{\centering\includegraphics[scale=0.35,trim={0.5cm 0.5cm 0.5cm 0.5cm},clip]{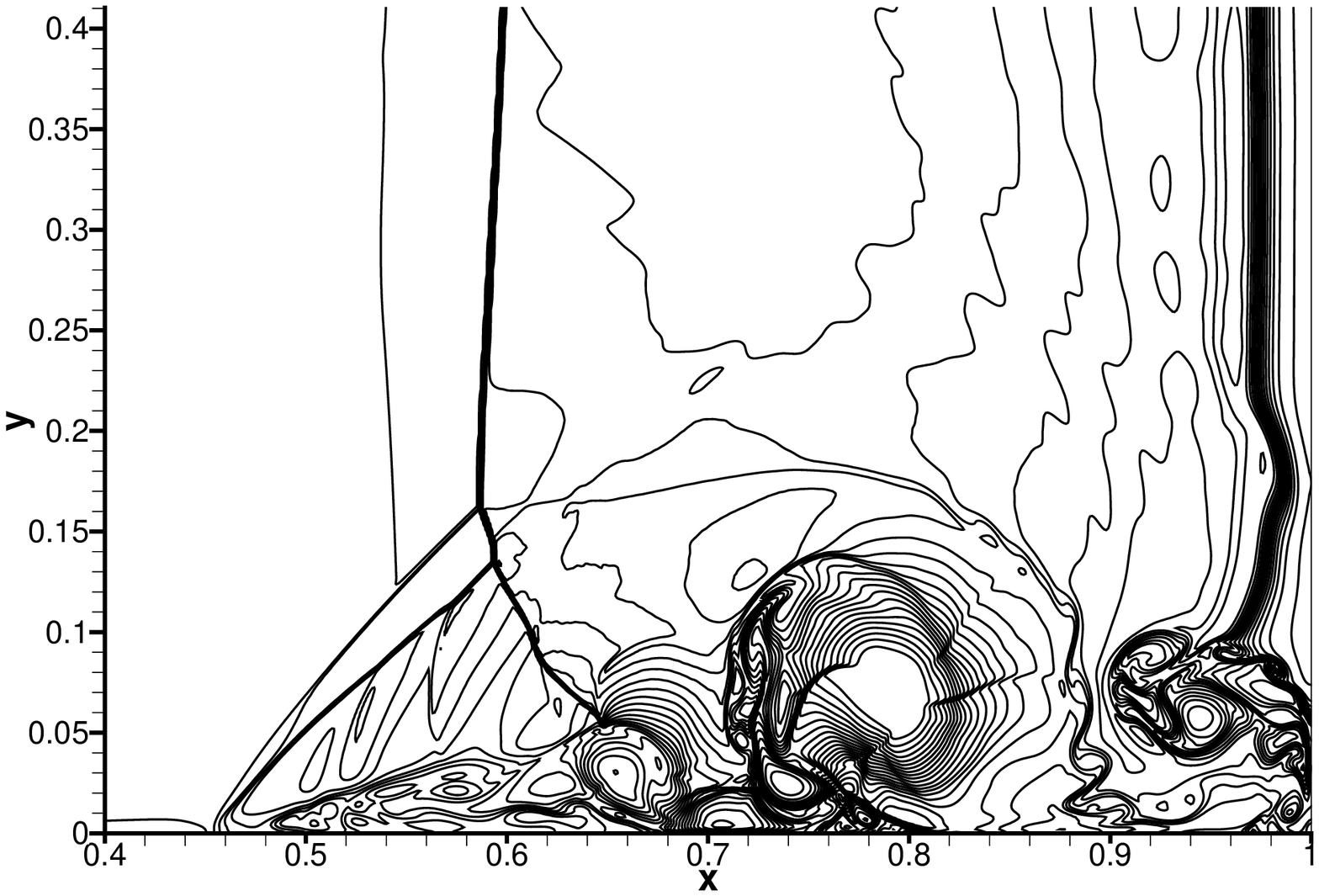}}
	\subfigure[9th order WENOM]{\centering\includegraphics[scale=0.35,trim={0.5cm 0.5cm 0.5cm 0.5cm},clip]{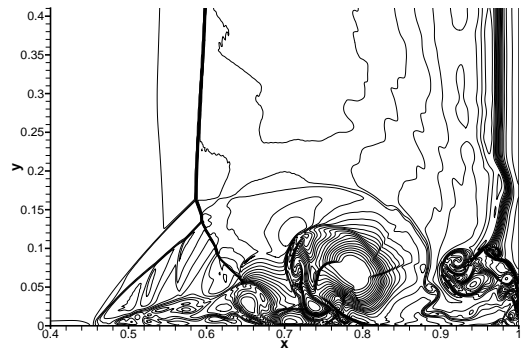}} 
    \end{center}
	\protect\caption{Simulation results at $t=1.0$ for 2D viscous shock tube with Re=1000. 21 contour levels between 20 and 120 are plotted. Comparisons were made between the $\mathrm{P}_{n}\mathrm{T}_{m}-\mathrm{BVD-CD}$ and high order WENO scheme. The computation was conducted with a mesh spacing of $\Delta x= \Delta y=0.001$.
		\label{fig:ShockBoundary}}	
\end{figure}

In this test, we employ 2D viscous shock tube to verify the performance of the proposed scheme in simulating unsteady viscous flows. The initial condition is set in a computational domain of $[0,1]\times[0,0.5]$ as 
\begin{equation}
(\rho_{0},\ u_{0},\ v_0,\ p_{0})=\left\{
\begin{array}{ll}
\left(120,\ 0,\ 0, \ \frac{120}{\gamma} \right), \ &\mathrm{if}\  x \leq 0.5,\\
\left(1.2,\ 0,\ 0, \ \frac{1.2}{\gamma} \right), \ &\mathrm{otherwise}.
\end{array}\right.
\end{equation}
The Prandtl number is $Pr=0.72$ and Reynolds number is $Re=1000$. The computation was conducted with a mesh spacing of $\Delta x= \Delta y=0.001$. A symmetric boundary condition is used on the top boundary. Non-slip adiabatic boundary is imposed at solid wall. A right-moving Mach 2.37 shock followed by a contact discontinuity will be created with this shock tube condition. The interaction between shock and the horizontal wall generates a thin boundary layer. After shock reflection on the right wall, the solution will develop complex interaction among shock, shear layer and boundary layer. The simulation results of density contour are presented in the Fig.~\ref{fig:ShockBoundary}. The result computed by the high order WENOM scheme is also included for comparison purpose. The simulation results show that the proposed $\mathrm{P}_{n}\mathrm{T}_{m}-\mathrm{BVD-CD}$ schemes resolve the delicate structures produced by interaction between shock and boundary layer. The resolved structure of the primary vortex agrees well with the DNS result shown in the Fig.~5(a) of \cite{ShockBoundaryDNS}. However, the primary vortex solved by the high order WENOM scheme is similar to the diffusive result in the Fig.~5(d) of \cite{ShockBoundaryDNS}. Compared with upwind-biased high order WENOM scheme, the $\mathrm{P}_{n}\mathrm{T}_{m}-\mathrm{BVD-CD}$ schemes are more preferable for DNS of compressible turbulence flow.       

\subsection{ILES of compressible isotropic turbulence}
In this subsection, we simulate the decaying compressible isotropic turbulence with eddy shocklets \cite{8} through the ILES approach. In this problem, if a sufficiently high turbulent Mach number is provided weak shock waves will develop from the turbulent motions. Thus the coexist of shock waves and turbulence requires the shock-capturing scheme should be able to simultaneously handle shocklets and broadband turbulence motion. The computational domain is a cubic volume of side length $2\pi L$ with periodic boundary condition. The initial condition consists of a random isotropic velocity field velocity fluctuations satisfying a prescribed energy spectrum as
\begin{equation}
E_{k} =u_{rms,0}^{2}16\sqrt{\dfrac{2}{\pi}}\dfrac{k^{4}}{k_{0}^{5}}\text{exp}(-\dfrac{2k^{2}}{k_{0}^{2}})
\end{equation}
where $u_{rms,0}$ is the root mean square turbulence intensity and the wavenumber $k_{0}=4$. The initial turbulent Mach number and Taylor-scale Reynolds number are defined as
\begin{equation}
M_{t,0}=\sqrt{3}u_{rms,0}/c_{0},\ \ 
Re_{\sigma,0}=\rho_{0}u_{rms,0}\sigma_{0}/\mu_{0}
\end{equation}
where $\sigma_{0}=\dfrac{2}{k_{0}}$ and $\mu_{0}$ is initial dynamic viscosity based on the initial temperature $T_{0}$. The power law is used to calculate the viscosity. Initially, the constant density and pressure are prescribed with $M_{t,0}=0.6$ and $Re_{\sigma,0}=100$. The problem is solved till a final time of $t/\tau=4$ where $\tau=\sigma/u_{rms,0}$ is the turbulent time scale. 

\begin{figure}
	\begin{center}
	\subfigure[$\mathrm{P_{4}T_{2}}$-BVD-CD]{\centering\includegraphics[width=.32\textwidth,trim={0.5cm 0.5cm 0.5cm 0.5cm},clip]{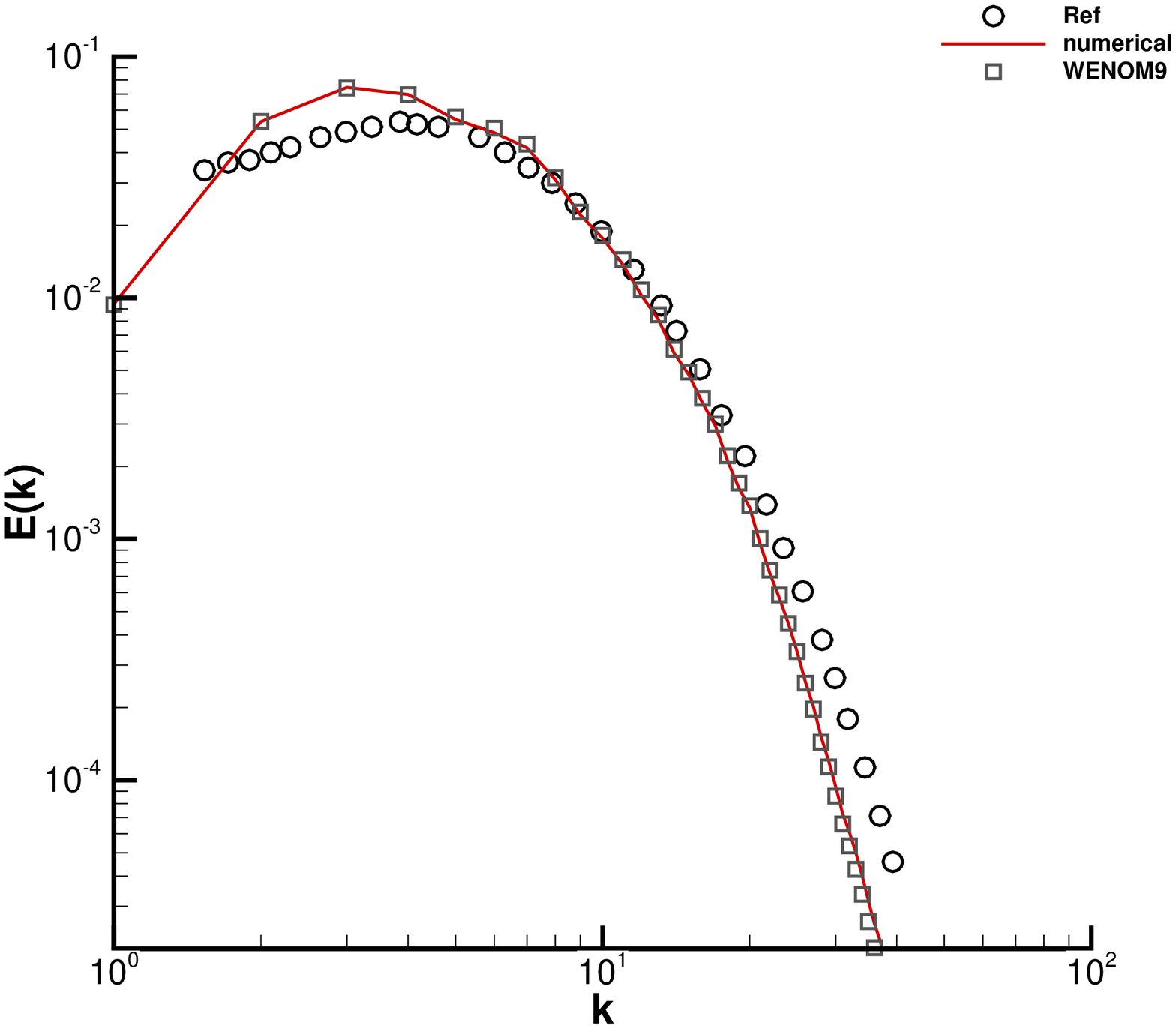}} 
	\subfigure[$\mathrm{P_{6}T_{3}}$-BVD-CD]{\centering\includegraphics[width=.32\textwidth,trim={0.5cm 0.5cm 0.5cm 0.5cm},clip]{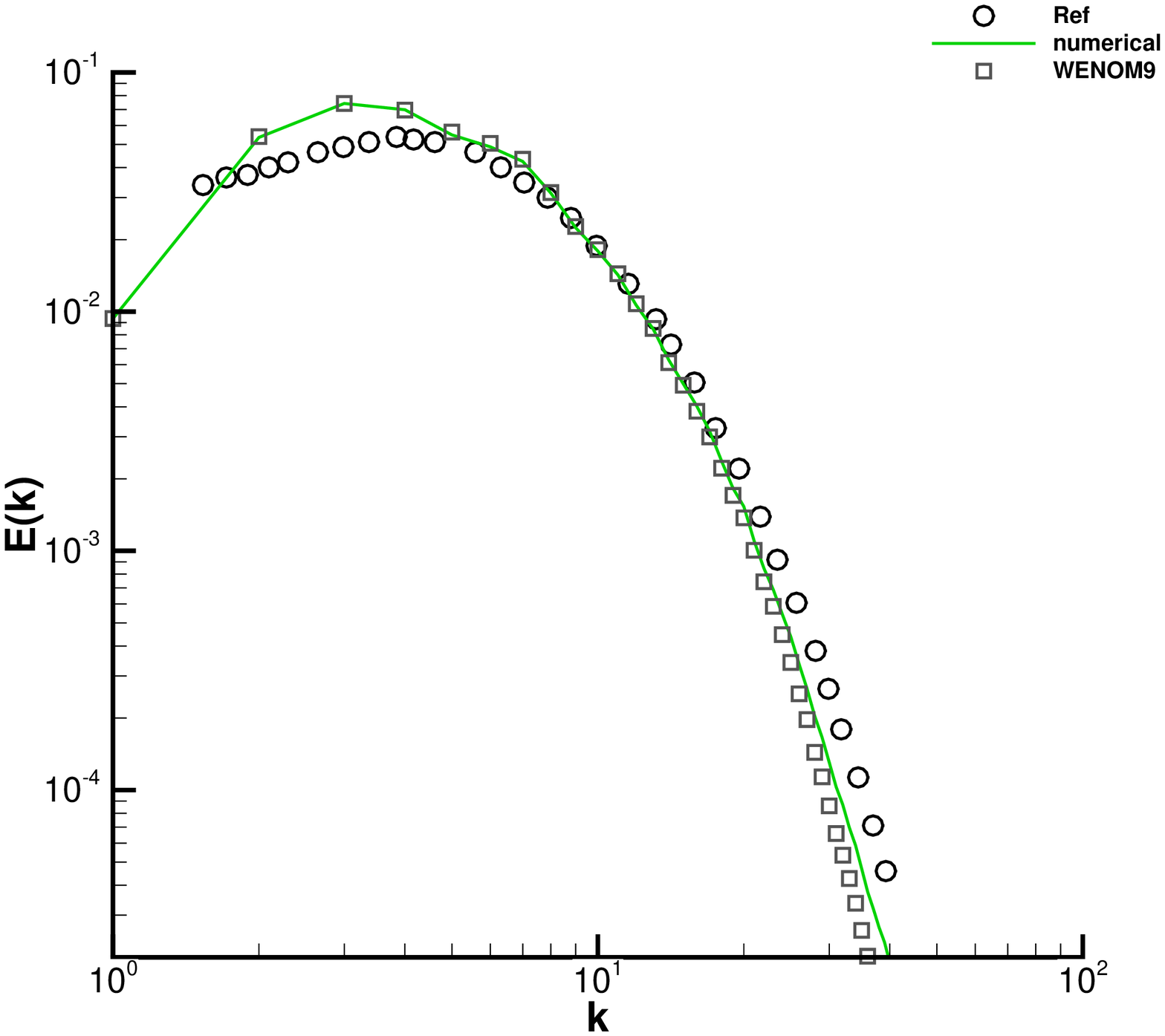}} 
	\subfigure[$\mathrm{P_{8}T_{3}}$-BVD-CD]{\centering\includegraphics[width=.32\textwidth,trim={0.5cm 0.5cm 0.5cm 0.5cm},clip]{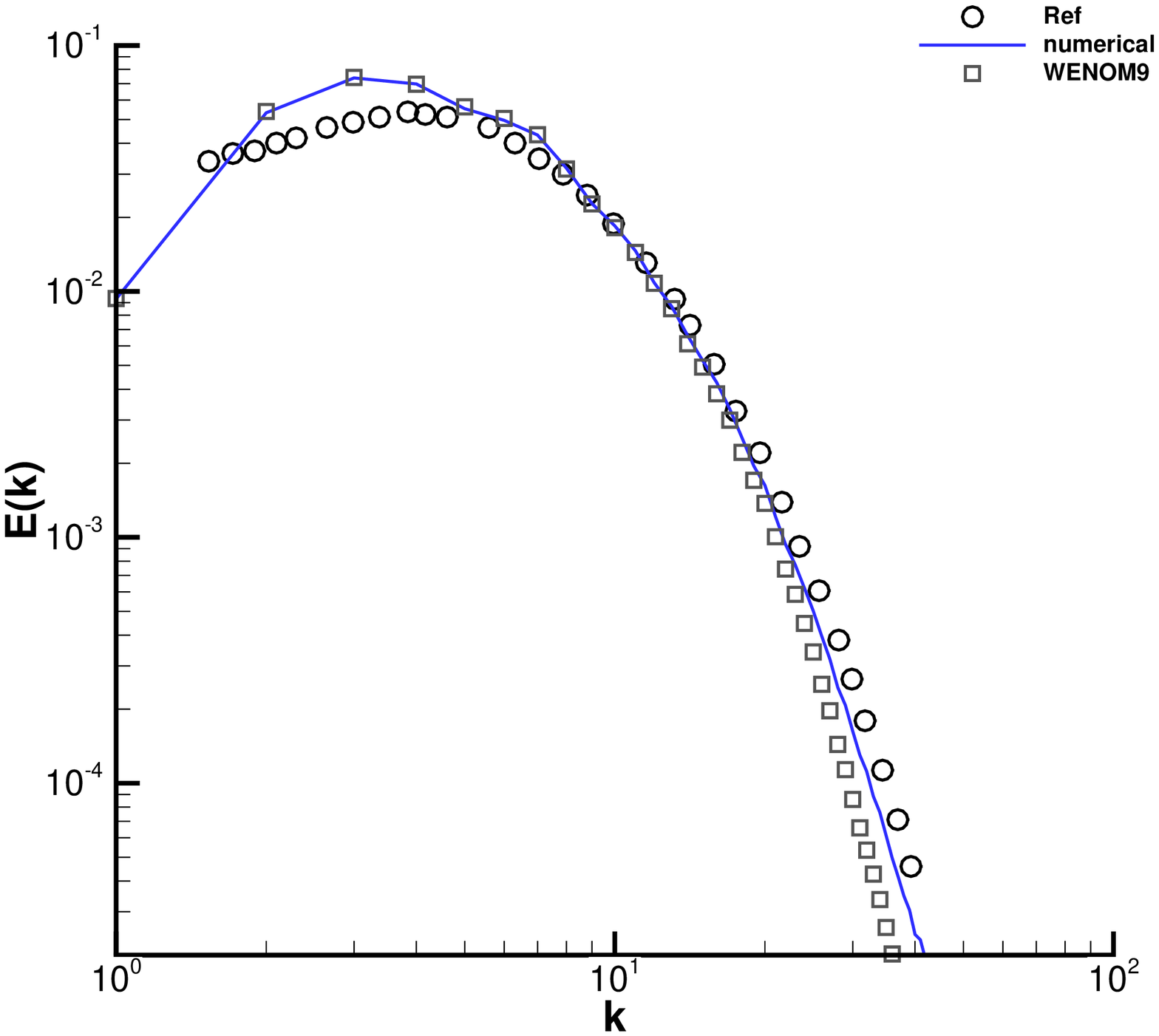}} 
    \end{center}
	\protect\caption{ILES of compressible isotropic turbulence. Velocity spectra at $t/\pi=4$ is presented. Comparisons were made between $\mathrm{P}_{n}\mathrm{T}_{m}-\mathrm{BVD-CD}$ and WENOM schemes on $64^{3}$ grids. The $\lambda=0.75$ is used for under-resolved ILES.
    \label{fig:hitSpectral}}	
\end{figure}

We first calculate the decaying compressible isotropic turbulence with a coarse mesh of $64^3$ cells. With the coarse mesh, the turbulent model such as LES is required to simulate the under-resolved turbulent structure. In our simulation, the dissipation controllable $\mathrm{P}_{n}\mathrm{T}_{m}-\mathrm{BVD-CD}$ schemes with $\lambda=0.75$ are employed to conduct ILES. The additional dissipation is used for ILES purpose. The numerical results for the velocity spectrum at $t/\tau=4$ are presented in Fig.~\ref{fig:hitSpectral} where the ILES conducted by ninth order WENOM scheme and the DNS result from \cite{8} are also included. It is obvious that $\mathrm{P}_{n}\mathrm{T}_{m}-\mathrm{BVD-CD}$ schemes of different order produce more accurate results than the ninth order WENOM scheme. As order increasing, $\mathrm{P}_{8}\mathrm{T}_{3}-\mathrm{BVD-CD}$ obtains results close to the DNS result. Thus, the numerical dissipation introduced by a tunable $\lambda$ can be employed to model the under-resolved scale through ILES. Compared with ninth order WENOM, the proposed $\mathrm{P}_{n}\mathrm{T}_{m}-\mathrm{BVD-CD}$ obtains more accurate results. More importantly, the numerical dissipation of the new scheme can be controlled more effectively, which allows investigation of adaptive control to improve ILES results in our future work. 

\begin{figure}
	\begin{center}
	\subfigure[$\mathrm{P_{4}T_{2}}$-BVD-CD]{\centering\includegraphics[width=.32\textwidth,trim={0.5cm 0.5cm 0.5cm 0.5cm},clip]{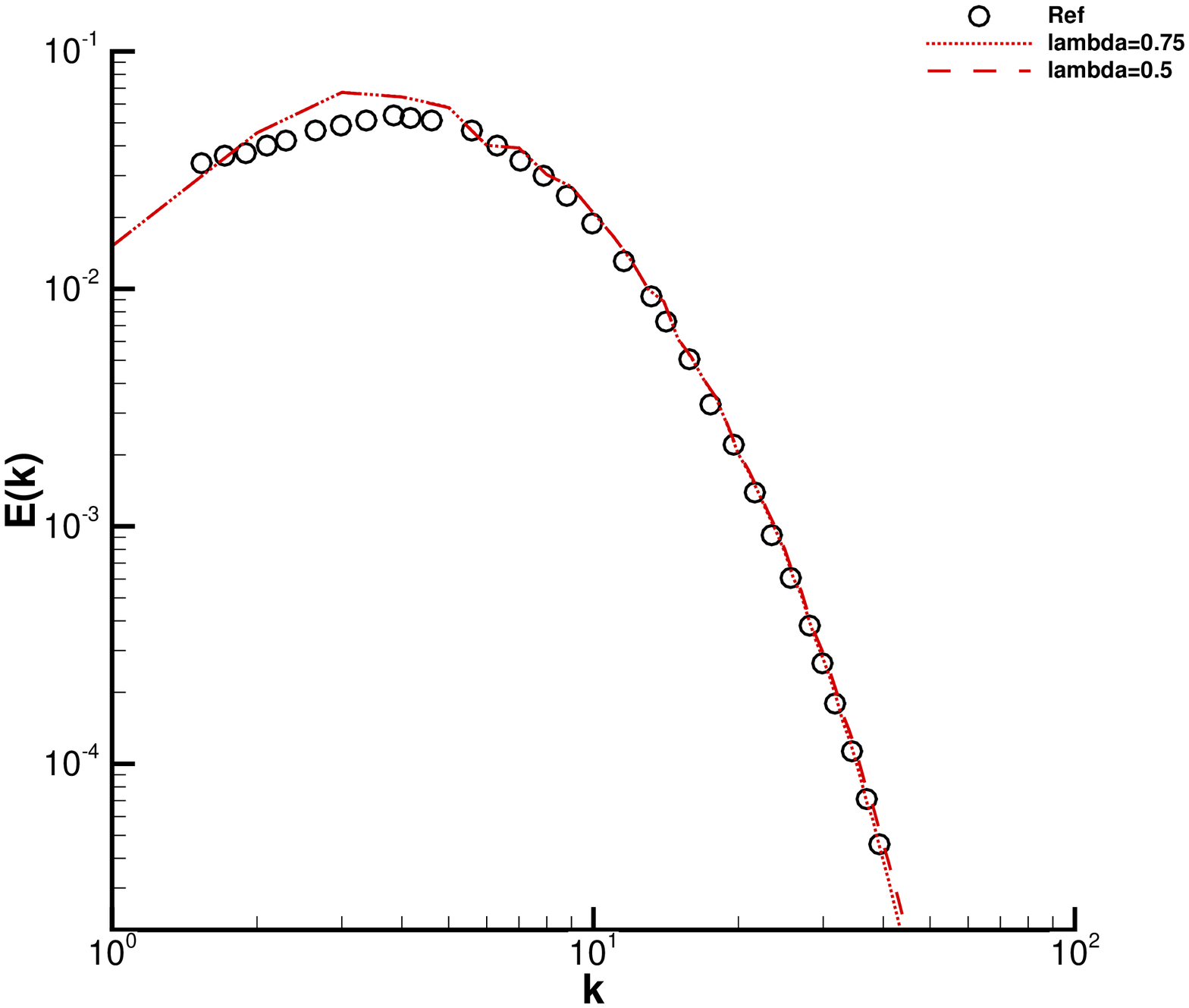}} 
	\subfigure[$\mathrm{P_{6}T_{3}}$-BVD-CD]{\centering\includegraphics[width=.32\textwidth,trim={0.5cm 0.5cm 0.5cm 0.5cm},clip]{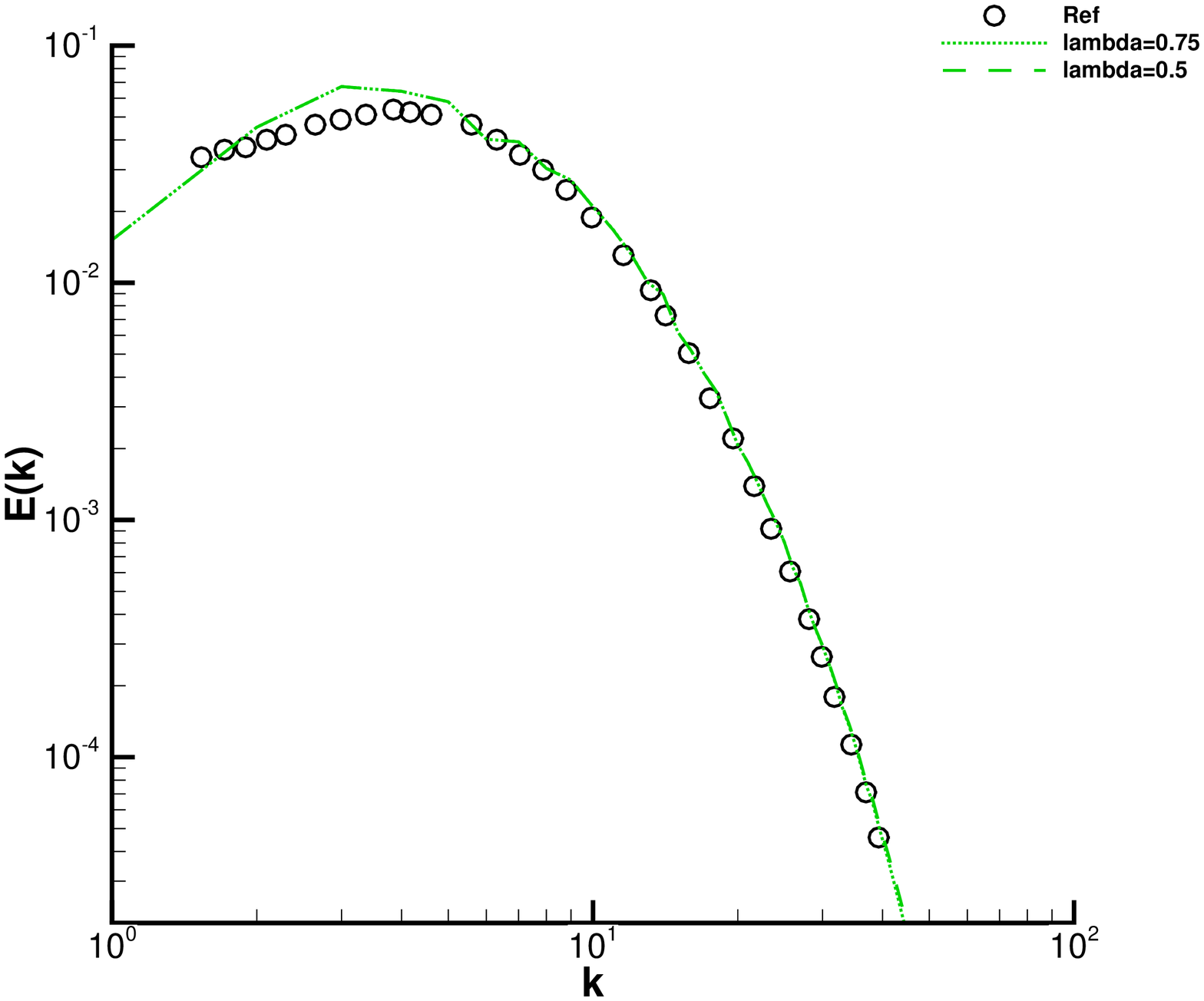}} 
	\subfigure[$\mathrm{P_{8}T_{3}}$-BVD-CD]{\centering\includegraphics[width=.32\textwidth,trim={0.5cm 0.5cm 0.5cm 0.5cm},clip]{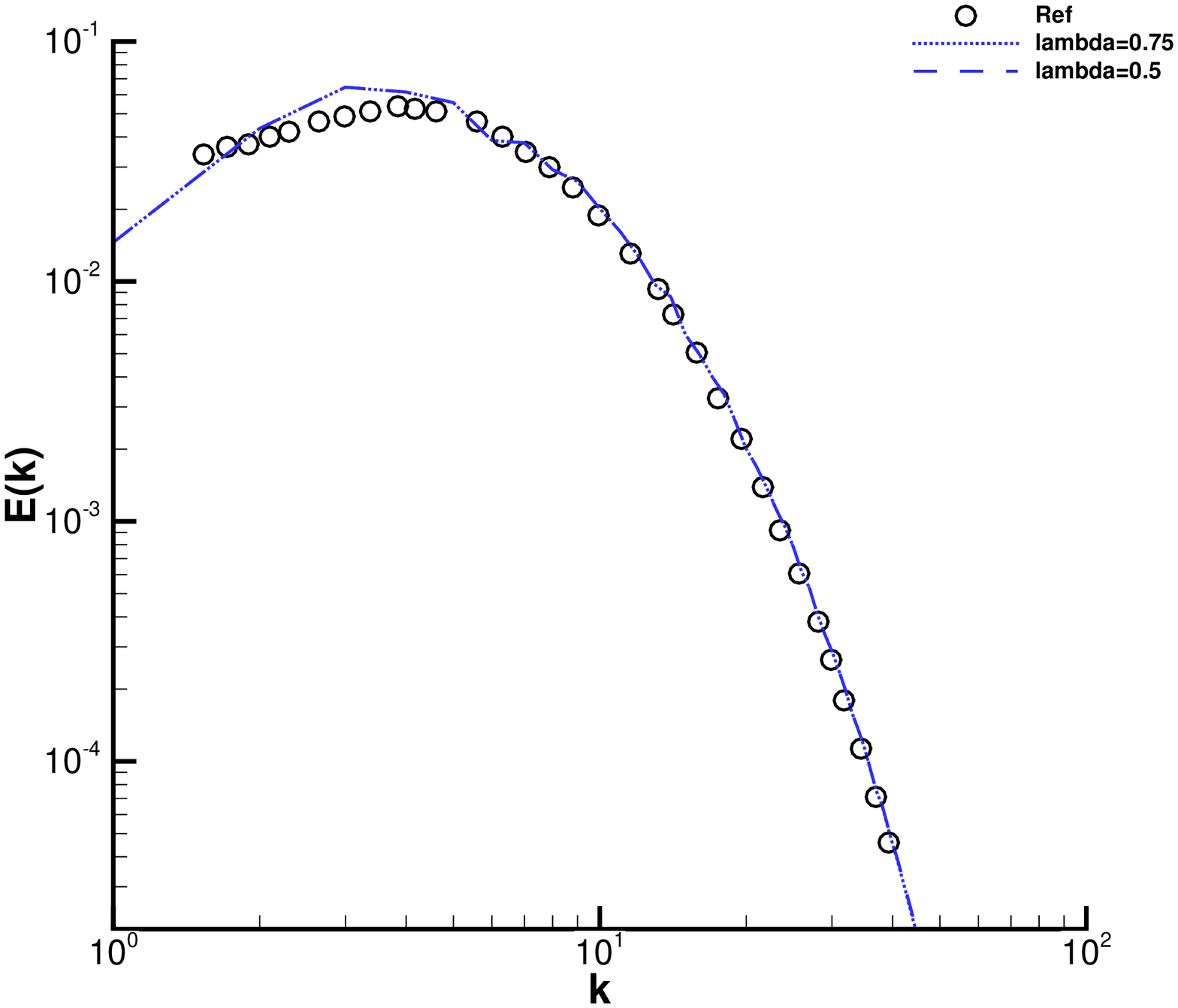}} 
    \end{center}
	\protect\caption{Convergence test for compressible isotropic turbulence problem with $\mathrm{P}_{n}\mathrm{T}_{m}-\mathrm{BVD-CD}$ using different $\lambda$. $128^{3}$ grid elements are employed. The DNS result marked by the circle is calculated with $256^{3}$ from \cite{8}. 
    \label{fig:hitSpectralCon}}	
\end{figure}

Finally, we conduct the convergence test for the compressible isotropic turbulence with different parameter $\lambda$ on the $128^3$ mesh elements. The numerical results regarding to the velocity spectrum are presented in the Fig.~\ref{fig:hitSpectralCon} where schemes with added dissipation through $\lambda=0.75$ and schemes with central flux through $\lambda=0.5$ are tested respectively. All of results converge to the DNS result, which should be expected since there are more resolvable scales with finer mesh elements. Thus the new $\mathrm{P}_{n}\mathrm{T}_{m}-\mathrm{BVD-CD}$ schemes provide faithful solutions for the LES and DNS of compressible turbulence flow.

\section{Concluding Remarks}
In this work, we propose a new paradigm of dissipation-controllable, multi-scale resolving scheme, named as PnTm-BVD-CD, for scale-resolving simulations. The desirable properties of the proposed schemes have been verified through benchmark tests involving multi-scale flow structures. Compared with high order upwind-biased WENOM schemes or other central-upwind WENO schemes, the PnTm-BVD-CD schemes resolve discontinuous large-scale structures with less numerical dissipation as well as less oscillation. With the boundary variation diminishing algorithm and the newly designed dissipation-controllable algorithm, the numerical dissipation of PnTm-BVD-CD can be effectively controlled between n+1 order upwind-biased scheme and non-dissipative n+2 order central scheme through a simple tunable parameter $\lambda$. The n+2 order non-dssipative central scheme can be retrieved over all wavenumber with $\lambda=0.5$, which is preferable for solving small-scale structures in DNS. Moreover, the under-resolved small-scale can be solved with the proposed scheme through the implicit LES approach. Thus this work proposes an accurate and robust scheme for scale-resolving simulation of high speed compressible flow involving broadband turbulence and strong shock waves.

Since the dissipation can be effectively controlled by the proposed scheme, we will explore the adaptive dissipation strategy inspired by the work \cite{autoDissipation} to improve the ILES approach in our future work. The proposed methodology will also be extended to finite difference schemes, compact schemes as well as be equipped with positivity-preserving algorithm such as \cite{raphael11,raphael12}.

\section*{Acknowledgment}
This work was supported in part by funds from Engineering and Physical Sciences Research Council (EP/R030340/1), and National Natural Sience Foundation of China (11702015).

%\bibliographystyle{elsarticle-num}
%\bibliography{references}

\clearpage{}

\end{document}